\documentclass[%
 reprint,
superscriptaddress,
 amsmath,amssymb,
 aps,
pra,
]{revtex4-2}

\usepackage{amsfonts}
\usepackage{amsmath}
\usepackage{stmaryrd}
\usepackage{graphicx}
\usepackage{dcolumn}
\usepackage{bm}
\usepackage[urlcolor=blue,hyperindex,colorlinks,bookmarks=true,linkcolor=black,citecolor=black]{hyperref}
\usepackage{physics}
\usepackage[english]{babel}

\usepackage{tcolorbox}
\usepackage{orcidlink}
\usepackage{mathrsfs}
\usepackage{tikz}
\usetikzlibrary{quantikz2}

\tcbuselibrary{theorems}

\newtcbtheorem[]{mytheo}{}%
{colback=blue!5,colframe=blue!50!black,fonttitle=\bfseries}{th}

\usepackage{tikz}
\usetikzlibrary{calc} 
\usetikzlibrary{positioning}
\usetikzlibrary{shapes.geometric}
\usetikzlibrary{quantikz2}
\usetikzlibrary{shadows}

\global\long\def\Tr{\operatorname{Tr}}%
\global\long\def\isdef{\coloneqq}%
\global\long\def\ket#1{\left|#1\right\rangle }%
\global\long\def\bra#1{\left\langle #1\right|}%
\global\long\def\braket#1#2{\left\langle #1\middle|#2\right\rangle }%
\global\long\def\ketbra#1#2{\left|#1\vphantom{#2}\right\rangle \left\langle \vphantom{#1}#2\right|}%
\global\long\def\kb#1#2{\left|#1\vphantom{#2}\right\rangle \left\langle \vphantom{#1}#2\right|}%
\global\long\def\braOket#1#2#3{\left\langle #1\middle|#2\middle|#3\right\rangle }%
\global\long\def\kk{\rangle\!\rangle}%
\global\long\def\bb{\langle\!\langle}%
\global\long\def\Op{\operatorname{Op}}%

\begin{document}

\title{Quantum Enhanced Pauli Propagation}

\author{S.~Majumder}
\email{swarnadeep.majumder@ibm.com}
\affiliation{IBM Quantum, IBM T.~J.~Watson Research Center, Yorktown Heights, NY 10598, USA}
\author{J.R.~Garrison}
\affiliation{IBM Quantum, IBM T.~J.~Watson Research Center, Yorktown Heights, NY 10598, USA}
\author{L. Luo}
\affiliation{IBM Quantum, IBM T.~J.~Watson Research Center, Yorktown Heights, NY 10598, USA}
\author{B.~Mitchell}
\affiliation{IBM Quantum, IBM Almaden Research Center, San Jose, CA 95120, USA}
\author{M.~Amico}
\affiliation{IBM Quantum, IBM T.~J.~Watson Research Center, Yorktown Heights, NY 10598, USA}
\author{A.~Seif}
\affiliation{IBM Quantum, IBM T.~J.~Watson Research Center, Yorktown Heights, NY 10598, USA}
\author{M.~Tran}
\affiliation{IBM Quantum, IBM T.~J.~Watson Research Center, Yorktown Heights, NY 10598, USA}
\author{K.~Sharma}
\affiliation{IBM Quantum, IBM T.~J.~Watson Research Center, Yorktown Heights, NY 10598, USA}
\author{E.~van den Berg}
\affiliation{IBM Quantum, IBM T.~J.~Watson Research Center, Yorktown Heights, NY 10598, USA}
\author{Z.~Minev}
\altaffiliation{Present address: Google Quantum AI, }
\affiliation{IBM Quantum, IBM T.~J.~Watson Research Center, Yorktown Heights, NY 10598, USA}

\author{L.~C.~G.~Govia}
\email{lcggovia@gmail.com}
\altaffiliation{Present address: CMC Microsystems, Waterloo, ON, Canada.}
\affiliation{IBM Quantum, IBM Almaden Research Center, San Jose, CA 95120, USA}

\date{\today}

\begin{abstract}

Accurately estimating observables on noisy quantum devices remains a central challenge for near‑term quantum algorithms. While quantum error mitigation techniques can reduce noise-induced bias, they often rely on unverifiable assumptions about the circuit noise, and cannot guarantee the magnitude of residual bias error. Here, rather than using classical resources to mitigate a noisy quantum circuit execution, we propose a hybrid algorithm that uses quantum resources to improve the accuracy of approximate classical Pauli‑path simulation. Our protocol, Quantum Enhanced Pauli Propagation (QuEPP), uses Clifford perturbation theory (CPT) to construct a classically simulable ensemble of Clifford circuits from the low-order terms in CPT, which directly provide the approximate classical Pauli-path simulation of the target circuit. Noisy quantum expectation values of this ensemble are then used to infer a global rescaling factor that corrects quantum execution of the target circuit, providing higher‑order contributions absent from the truncated low-order classical simulation. This approach requires no noise characterization, applies to arbitrary circuits, and provides a provable route to asymptotically unbiased estimates. Using IBM Heron hardware, we demonstrate QuEPP on 2D random mirror circuits of up to 49 qubits and circuit depth 80, as well as Trotterized Hamiltonian evolution, showing consistent improvements beyond classical CPT and unmitigated quantum results. QuEPP offers a simple, scalable, and model‑free framework for enabling accurate quantum computation in the pre‑fault‑tolerant era.

\end{abstract}

\maketitle

\section{Introduction} 

With the advent of utility-scale quantum computers \cite{utility}, quantum error mitigation (QEM) has attracted great attention as a building block towards quantum advantage on near-term systems. Broadly speaking, all QEM protocols leverage additional quantum circuit executions and classical post-processing to undo the effect of error in the noisy execution of a target circuit \cite{Cai23}. As the resources required for this classical post-processing typically scale exponentially with the underlying noise, the emergence of QEM has brought about renewed interest in (exponentially-scaling) classical simulation of quantum circuits. Contemporary methods such as tensor networks and Pauli propagation \cite{Begusic23,rudolph25} have proven to be competitive with utility-scale quantum computations for some circuits.

Both QEM and some classical methods leverage assumptions about the way error propagates through the target circuit. With rare exceptions \cite{Liu24,Wang25}, QEM protocols assume that the error is Markovian, and described by a completely-positive and trace-preserving channel. Some QEM protocols \cite{utility,Berg2023,Filippov23} further require a characterization of this error channel. To be scalable to large system sizes such characterizations must use an ansatz that assumes locality or other effective-dimension reducing restrictions on the error processes. Even when the true error in the circuit satisfies all assumptions of the protocol, for arbitrary target circuits many popular error mitigation protocols cannot produce an estimate to the expectation value that is bias free \cite{Czarnik2021,Farrell24_1,Li17,Endo18}. QEM protocols deploying detailed error characterization can overcome this limit \cite{Berg2023}, but in practice even these will have residual bias, for example due to finite accuracy of characterization \cite{Govia25,Filippov24} or noise drift \cite{kim2024}. Bounding bias is crucial to establish confidence in any computational method, but this has proven to be challenging for QEM, and heuristics based on classically tractable surrogate circuits are often deployed \cite{Govia25,Merkel25}.

In this work, rather than using classical resources to improve the accuracy of a quantum execution as in a QEM protocol, we use quantum resources to enhance the classical Pauli propagation method. Our protocol, which we dub Quantum Enhanced Pauli Propagation (QuEPP) returns an estimate to a circuit expectation value with a bounded bias error that can be systematically reduced to zero. It does not require error characterization, and makes no restrictions on circuit structure or locality of circuit error. QuEPP deploys a simple error mitigation strategy for the quantum executions, and achieves near optimal variance scaling in terms of circuit noise \cite{Regula2021,Takagi22}. As the error mitigation is performed entirely in classical post-processing, by design QuEPP naturally fits into the quantum combined with High Performance Computing paradigm.

This paper is organized as follows. In section \ref{sec:PP} we review the Pauli propagation method to motivate the QuEPP protocol, which we describe in section \ref{sec:queep}. In section \ref{sec:exp} we present experimental results, on up to 49 qubits and circuit depth 80, which demonstrate that QuEPP can be deployed accurately at scale. 

Finally, in section \ref{sec:conc} we make general and concluding remarks.

\section{Pauli Propagation and Pauli Paths}
\label{sec:PP}
Pauli propagation is a technique for estimating expectation values by evolving observables backward through a circuit in the Heisenberg picture. Instead of simulating the state forward, we express the observable $O$ in the Pauli basis and compute
\[
\langle O \rangle = \bra{\psi} U^\dagger O U \ket{\psi},
\]
where $U$ is the circuit and $\ket{\psi}$ the input state. Each Pauli term is conjugated gate by gate: Clifford gates map Paulis to Paulis deterministically, while non-Clifford gates map a Pauli to a \emph{linear combination} of Paulis, introducing branching. These branches define \emph{Pauli paths}, and the weighted sum of the expectation values over all paths gives the exact observable value. This perspective enables efficient simulation for near-Clifford circuits, and forms the basis for Monte Carlo sampling and truncation-based algorithms for approximate simulation of general circuits.

For example,  consider the single-qubit circuit $U = R_x(\theta) H$, where $H$ is the Hadamard gate (Clifford) and $R_x(\theta) = e^{-i\theta X/2}$ is a rotation about $X$ (non-Clifford for generic $\theta$). For output observable $Z$, we propagate backward
\[
R_x(\theta)^\dagger Z R_x(\theta) =  \cos\theta Z +  \sin\theta Y,
\]
so the non-Clifford gate splits $Z$ into two Pauli terms. Through $H$, these evolve as
\[
H^\dagger Z H = X, \qquad H^\dagger Y H = -Y,
\]
giving
\[
U^\dagger Z U = \cos\theta\, X - \sin\theta\, Y.
\]
Thus, two Pauli paths emerge: $(Z \to X)$ with weight $\cos\theta$ and $(Z \to Y)$ with weight $-\sin\theta$. The exact expectation is
\[
\langle Z \rangle = \cos\theta\, \bra{\psi} X \ket{\psi} - \sin\theta\, \bra{\psi} Y \ket{\psi},
\]
illustrating that the target expectation value can be written as a sum over the different paths.

\begin{figure*}[t!]
\setlength{\tabcolsep}{10pt}
\centering
\includegraphics[width=1\textwidth]{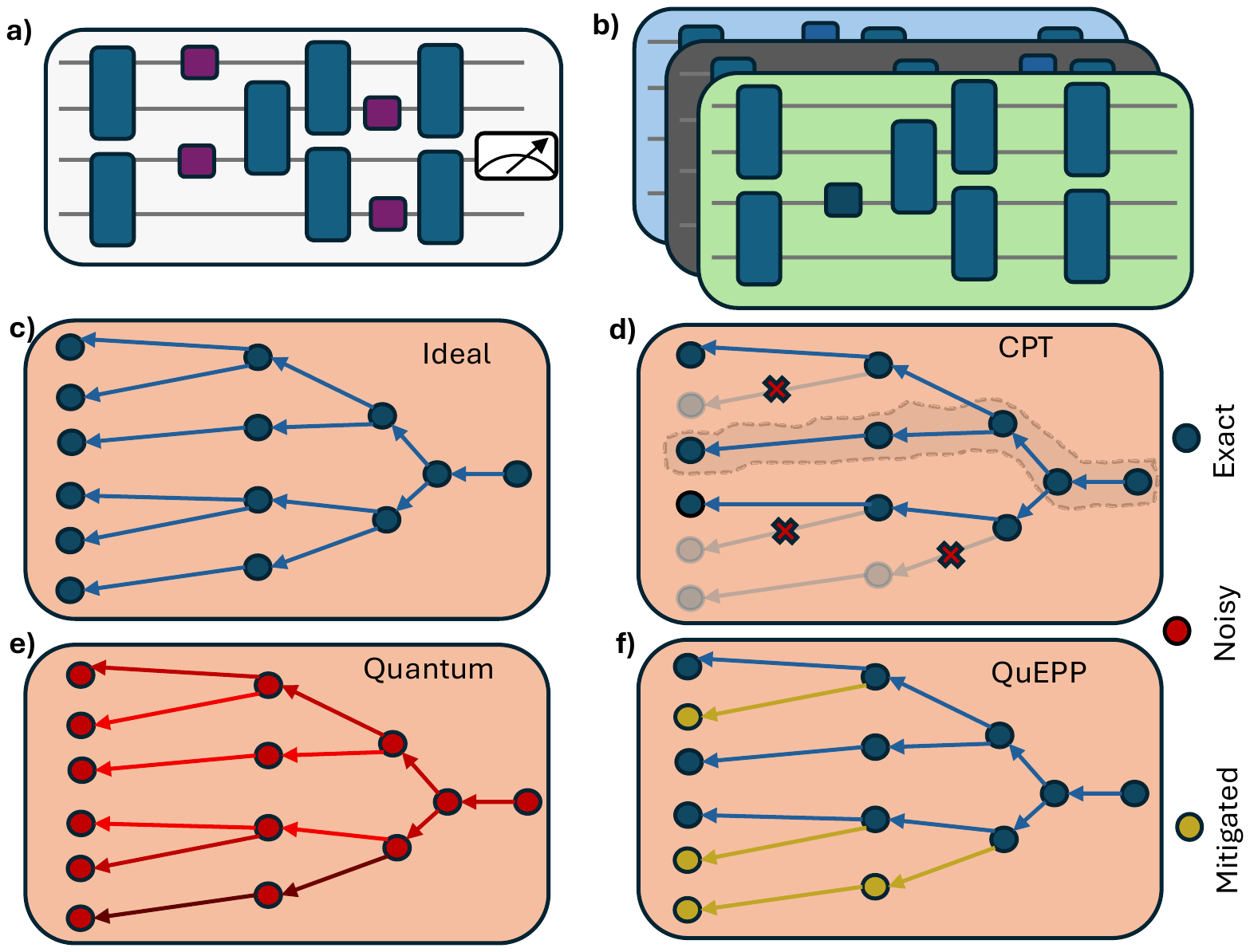}
\caption{\textbf{Quantum Enhanced Pauli Path Simulation with QuEPP.} a) The input to QuEPP is a quantum circuit and observable. The circuit consists of Clifford gates (blue), Pauli rotations (purple), and measurement. (b) Clifford Perturbation Theory (CPT) can be used to write the target expectation value as a sum over expectation values of many Clifford circuits. (c) This sum can be represented by Pauli paths describing how the observable is propagated through the circuit in a tree like graph. Here, nodes are Cliffords or Pauli rotations that compose the evolution of the observable, and edges indicate propagation of the observable along Pauli paths; the diagram shows path structure, not literal circuit edges. Clifford operations map a Pauli to a different Pauli, whereas Pauli rotations can split the Pauli into two different paths. (d) CPT truncates some of these paths based on the order to remain computationally feasible. Each path maps to a single Clifford circuit. (e) Running the target circuit directly gives us a noisy estimate of all the paths combined, and running the CPT ensemble Clifford circuits of the lower order terms gives the noisy estimate of the lower order paths. Different shades of red colors are used to denote the noisy paths. (f) Combining exact CPT estimate of lower order terms with rescaled estimate of the higher order terms, we get a better estimate of the expectation value than (d) or (e) alone. 
}
\label{fig1:Pauli Path}
\end{figure*}

\begin{figure*}[t!]
\setlength{\tabcolsep}{10pt}
\centering
\includegraphics[width=1\textwidth]{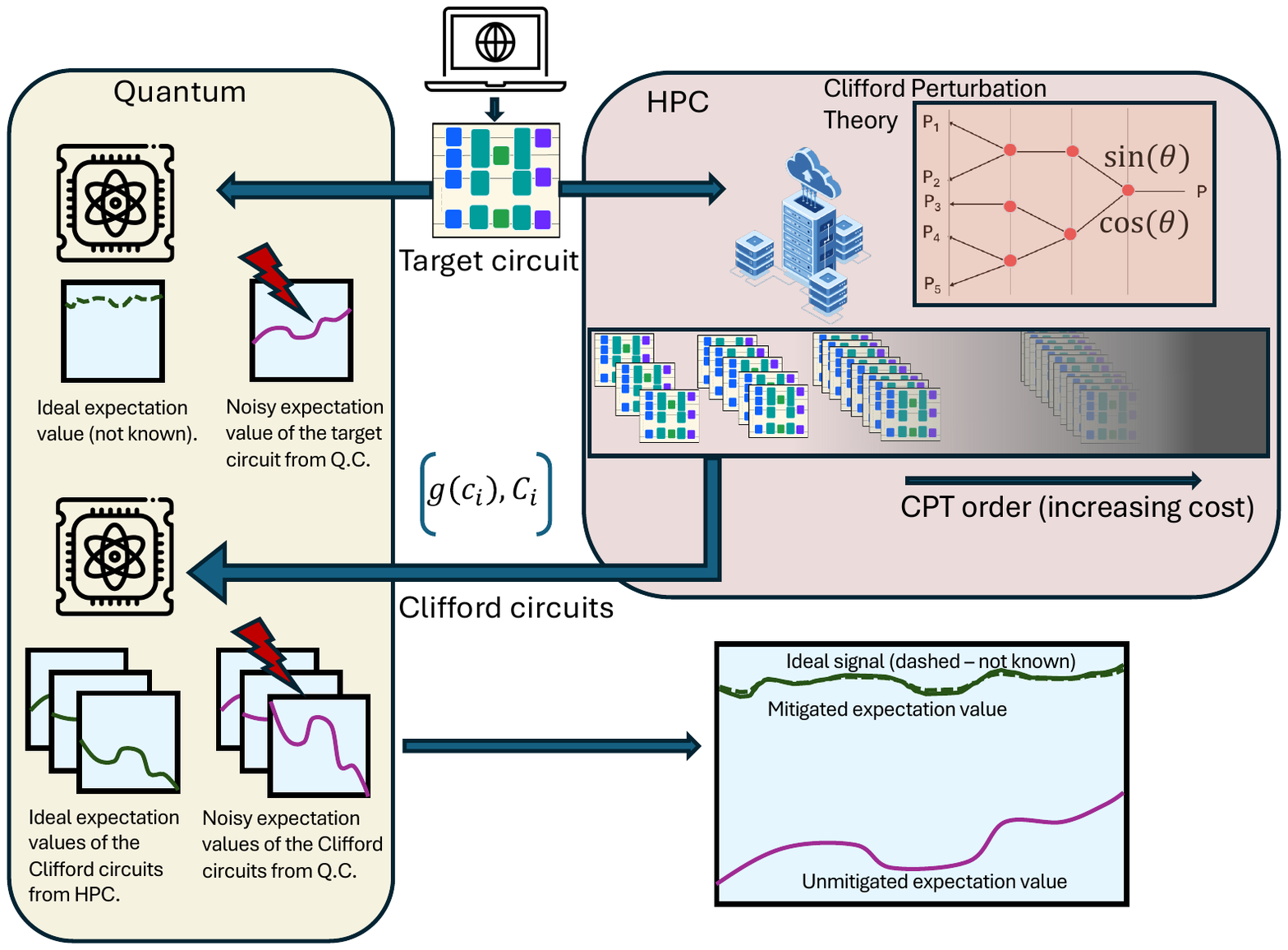}
\caption{\textbf{Mode of Operation for QuEPP} This figure complements Fig.\ \ref{fig1:Pauli Path} by showing the operation workflow rather than the underlying Pauli-path structure. The user begins by specifying both a target quantum circuit and an observable on a classical computer. This circuit is sent to both a quantum computer for execution and an HPC cluster for further processing. The quantum computer collects shots for this target circuit to estimate the noisy expectation value of the observable. Meanwhile, the HPC cluster performs CPT up to a fixed order and collects Clifford circuits to form the CPT ensemble. Each of these Clifford circuits that has a non-zero overlap with the intended initial state is sent to be also run on the quantum computer. The noisy and exact expectation values of the Clifford circuits are used to estimate the impact of noise with QuEPP, and this correction is used to result in a mitigated expectation value of the observable with respect to the original (target) circuit.   
}
\label{fig2:Mode of Operation}
\end{figure*}

\section{The Quantum Enhanced Pauli Propagation Protocol}
\label{sec:queep}

The starting point for QuEPP is the Pauli propagation protocol known as Clifford perturbation theory (CPT) \cite{Begusic23}, which reformulates any circuit into a transparent structure for Pauli propagation, with alternating layers of Clifford gates and non-Clifford Pauli rotation gates $\exp\left(-i\theta P/2\right)$. The expectation value of an observable $O$ and initial state $\rho$ propagated through a target circuit $\mathcal{U}$ (in superoperator notation) can then be written as 
\begin{align}
    \left<O\right> = {\rm Tr}\left[\rho\mathcal{U}^\dagger(O)\right] = \sum_{k=0}^K \sum_{i=1}^{N_k} g(i,k){\rm Tr}\left[\rho\mathcal{C}_{i,k}^\dagger(O)\right], \label{eqn:cpt}
\end{align}
where $\mathcal{C}_{i,k}$ is a Clifford circuit in the so-called CPT ensemble, $K$ is the total number of Pauli rotation gates in the target circuit, and $N_k$ is the number of circuits in the ensemble at each order $k$. As Clifford circuits are classically efficient to simulate, the hardness of simulating the target circuit reduces to simulating enough of the Clifford ensemble defined by CPT, and there is evidence that hardness for CPT maps to hardness for other simulation methods \cite{Dowling25}.

As with any Pauli path technique, the expectation value is written as a sum over the paths, each represented by the Clifford circuit $\mathcal{C}_{i,k}$. A Pauli path can be represented by a Clifford circuit because the Pauli evolution along any path deterministically maps Paulis to Paulis. To see this, consider first that any Clifford gates in the target circuit map Paulis to Paulis and are added directly to the path circuit $\mathcal{C}_{i,k}$ at their target circuit locations. Consider next a rotation gate $R_P(\theta)$ in the target circuit where branching occurs. On a Pauli $O$ this generates the evolution
\begin{align}
\nonumber &R_P(\theta)^\dagger O R_P(\theta) =  \cos\theta O +  i\sin\theta PO \\
&=\cos\theta O +  \sin\theta R_P\left(\frac{\pi}{2}\right)^\dagger O R_P\left(\frac{\pi}{2}\right),
\end{align}
where we have used the definition of the rotation gate itself in the last line to rewrite the operator product $iPO$ as unitary evolution. As can be seen, the cosine branch replaces the rotation gate with the identity gate as $O$ is unchanged, while the sine branch replaces the general Pauli rotation with $R_P(\pi/2)$, which for all $P$ is a Clifford gate. Thus, all gates in the circuit $\mathcal{C}_{i,k}$ are Clifford.

The coefficient
\begin{align}
    g(i,k) = \prod_{j=1}^k\sin{\theta_{i,j}}\prod_{j=k+1}^{K}\left(\cos{\theta_{i,j}}\right)^{s_{i,j}}, \label{eqn:CPTcoeff}
\end{align}
gives the weight of a specific Pauli path, with the angles $\theta_{i,j}$ defined by the Pauli rotation gates in the target circuit. The parameter $s_{i,k}$ is zero (one) if the back-propagated observable commutes (anti-commutes) with the generator of the Pauli rotation gate at a given layer. By factoring out Clifford gates from Pauli rotation gates of arbitrary angle, we have that $\abs{\theta_{i,j}}\leq \pi/4$, and so $\sin{\theta_{i,j}}$ naturally defines a small parameter for the series. Finite truncations of the first summation in the CPT series of Eq.~\eqref{eqn:cpt} can be treated as a perturbative summation in the number of $\sin{\theta_{i,j}}$ terms, indexed by the order $k$. For \emph{most} circuits there exists an order $k'$ beyond which the higher order parts of the series monotonically decrease in contribution to the expectation value.

Using this series description of the target circuit, QuEPP combines quantum and classical resources to enhance traditional Pauli path simulations (see Fig. \ref{fig1:Pauli Path}). For simplification, we will assume that our circuits are Pauli twirled \cite{Bennett96,Knill04,Kern05,Geller13,Wallman16}, as it reduces the impact of error on the Clifford circuits in the CPT ensemble to a rescaling of the ideal expectation value. QuEPP uses a simple QEM protocol based on rescaling all quantum circuit executions by a single parameter $\eta$ (similar in spirit to Refs.~\cite{Czarnik2021,Farrell24_1}). This rescaling parameter is estimated by comparing the result of the noisy quantum executions of the CPT ensemble circuit to their ideal values.

With this background in place, we can now define the QuEPP protocol (see Fig.\ \ref{fig2:Mode of Operation}):

\begin{enumerate}
    \item Using classical compute, calculate the ideal expectation values for the CPT ensemble circuits up to order $K_T$: ${\rm Tr}\left[\rho\mathcal{C}_{i,k}^\dagger(O)\right]$.
    And the classical estimate up to order $K_T$
    \begin{align}
        \left<O\right>^{{K_T}} = \sum_{k=0}^{K_T} \sum_{i=1}^{N_k} g(i,k){\rm Tr}\left[\rho{\mathcal{C}}_{i,k}^\dagger(O)\right].
    \end{align}

    \item On the quantum computer estimate the noisy expectation values of: 
    
    \begin{enumerate}
        \item The target circuit: $\left<O\right>_{\rm noisy} = {\rm Tr}\left[\rho\tilde{\mathcal{U}}^\dagger(O)\right]$.
        \item The Clifford circuits in the CPT ensemble up to order $K_T$ that have non-zero ideal expectation: ${\rm Tr}\left[\rho\tilde{\mathcal{C}}_{i,k}^\dagger(O)\right]$. The noisy estimate up to order $K_{T}$ of the target circuit is 
        \begin{align}
            \left<O\right>^{{K_T}}_{\rm noisy} = \sum_{k=0}^{K_T} \sum_{i=1}^{N_k} g(i,k){\rm Tr}\left[\rho\tilde{\mathcal{C}}_{i,k}^\dagger(O)\right].
        \end{align}
        
    \end{enumerate}

    \item Subtract the noisy estimate of the expectation values up to order $K_T$ from the noisy expectation value of the target circuit. This gives us a noisy estimate of the paths we did not compute classically
    \begin{align}
        \nonumber\left<O\right>^{\lnot{K_T}}_{\rm noisy} = \left<O\right>_{\rm noisy} -  \left<O\right>^{{K_T}}_{\rm noisy}.
    \end{align}
    
    \item From the distribution of per-circuit rescaling factors $\eta_{i,k} = {\rm Tr}\left[\rho\tilde{\mathcal{C}}_{i,k}^\dagger(O)\right]/{\rm Tr}\left[\rho\mathcal{C}_{i,k}^\dagger(O)\right]$ of the CPT ensemble, systematically determine a global rescaling factor $\eta$.

    \item Compute the enhanced expectation value
    \begin{align}
        \nonumber\left<O\right>^{K_T}_{\mathbb{M}} = 
        \left<O\right>^{{K_T}} +  \left<O\right>^{\lnot{K_T}}_{\rm noisy} /\eta   
    \end{align}

\end{enumerate}
In the above we have used $\tilde{\mathcal{U}}$ ($\tilde{\mathcal{C}})$ to denote the noisy version of the ideal target (Clifford) circuit $\mathcal{U}$ ($\mathcal{C}$). As we show in the supplementary material, compared to approximate classical simulation of the target circuit via CPT, our QuEPP protocol has an eventual guarantee of improved accuracy. This can be understood by the fact that in QuEPP the missing terms in the complete CPT sum are estimated by the mitigated noisy value $\left<O\right>^{\lnot{K_T}}_{\rm noisy}/\eta$. Similarly, QuEPP is asymptotically bias free, and for truncation order $K_T$ the remaining bias can be upper bounded (see supplementary material).

We highlight two properties of QuEPP that make it particularly advantageous for near-term quantum computing.  Firstly, since only the per-circuit noise factors $\eta_{i,k}$ are measured, QuEPP places no restrictions on the structure of the target circuit, e.g.,~that it has clearly defined ``easy'' and ``hard'' layers. Secondly, it is a model-free error mitigation procedure that requires no separate characterization phase. This is beneficial in situations where the noise channel is known to drift in time \cite{kim2024}, as all the quantum executions (step 2 of the protocol) can be run interleaved in a single batch. To that point, we emphasize that the error mitigation in QuEPP is performed entirely in classical post-processing. This gives us the freedom to explore different methods for calculating $\eta$ without additional quantum compute. While we cannot know what the best choice of $\eta$ is without a way to verify our results, comparing different $\eta$ also acts as a consistency check for the output expectation value. In this work we use the median of the scale-factor distribution $\{\eta_{i,k}\}$, and discuss other options in the supplementary material.

As described, the operation mode of QuEPP would be to classically simulate as many orders of the CPT ensemble as possible, and then via QuEPP obtain a more accurate result using additional quantum executions. To fairly count the resources of QuEPP compared to classical CPT simulation, we must consider that each additional quantum circuit execution used in QuEPP requires many shots to reduce the variance of its expectation value to an acceptable level. In the supplementary material, we derive that the total variance in QuEPP is bounded by $\sigma^2 \leq \gamma P_{K_T}/N$ for $N$ executions of each quantum circuit, with $\gamma = 1/\sqrt{\eta}$ the resource parameter introduced for probabilistic error cancellation \cite{Berg2023}. The pre-factor $P_{K_T}$ is the squared sum of CPT coefficients up to order $K_T$, and is always less than 1. 

An alternative to the order based expansion is the instead sample different Clifford paths with Monte Carlo sampling. While this approach loses much of the theoretical guarantee of convergence of the order based method, it can be advantageous when the CPT ensemble is very large. We discuss the specific details of the Monte Carlo method in the supplementary material. 

\section{Experimental Results}
\label{sec:exp}
In order to experimentally demonstrate the effectiveness of QuEPP, we perform three experiments on IBM's latest generation of Heron processors. These experiments are of circuits whose ideal outcome is verifiable, either by construction (mirror circuits), or brute force classical simulation.

\subsection{Random Mirror Circuit}
\subsubsection{Easy-to-simulate regime}
\begin{figure*}[t!]
\setlength{\tabcolsep}{10pt}
\centering
\includegraphics[width=1\textwidth]{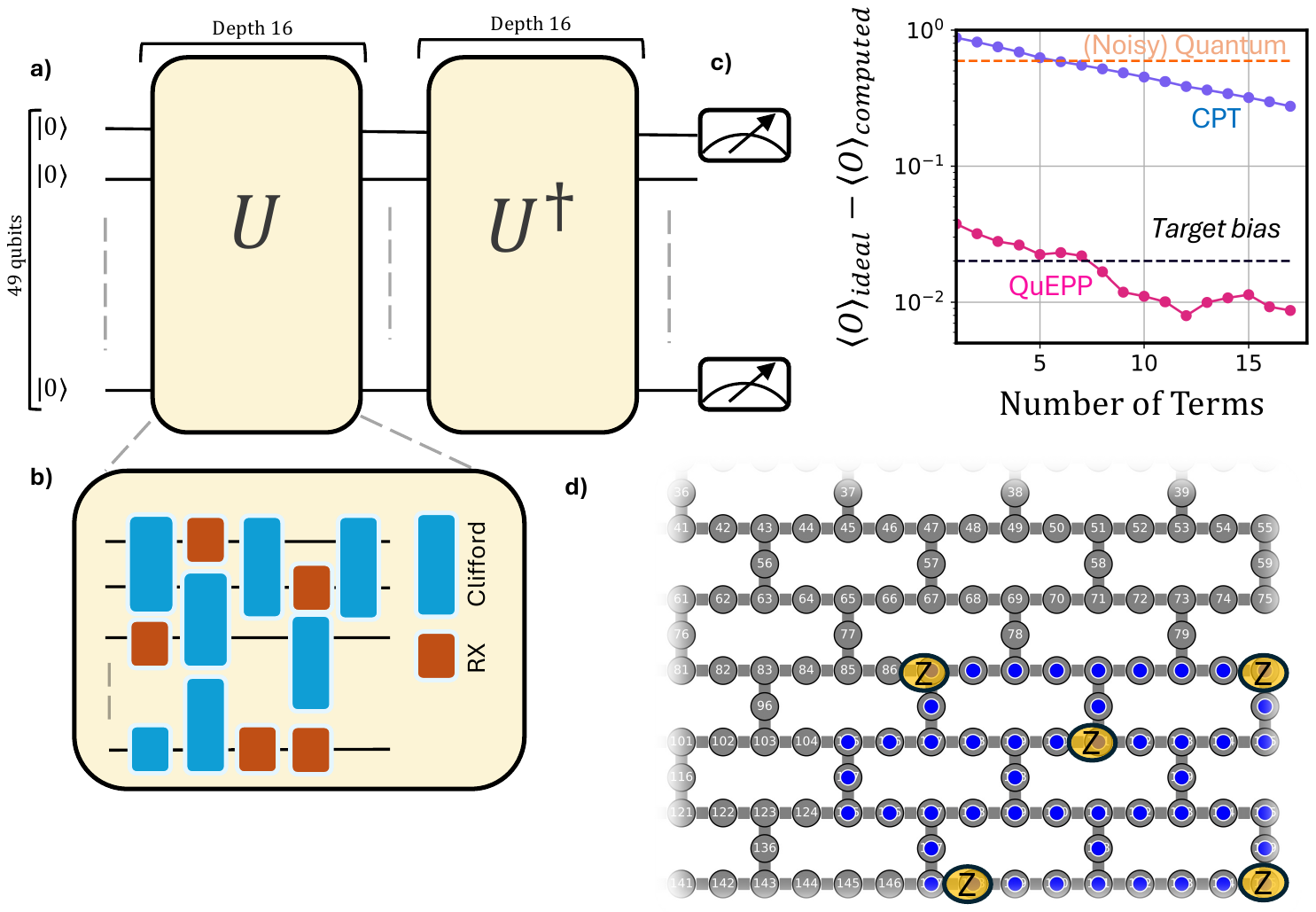}
\caption{\textbf{Unstructured mirror circuit}
a) The target circuit consists of preparing the all zero state on 49 qubits and 16 layers of a forward propagating random 2D circuit  followed by the mirror inverse circuit. We are interested in measuring the expectation value of the observable $\langle Z_{0}Z_{11}Z_{18}Z_{41}Z_{48} \rangle$. b) Layers in the circuit consists of single qubit and two qubit Clifford gates with some non-Clifford Pauli RX rotations.  c) Bias (as calculated by difference between ideal and computed expectation value) is plotted as a function of CPT order. We are comparing the purely classical method with CPT (in Purple) and QuEPP (in Pink). d) The circuit is mapped onto a heavy-hex topology while measured qubits are marked with yellow circles.  } 
\label{fig3:Random Mirror}
\end{figure*}

\begin{figure*}[t!]
\setlength{\tabcolsep}{10pt}
\centering
\includegraphics[width=1\textwidth]{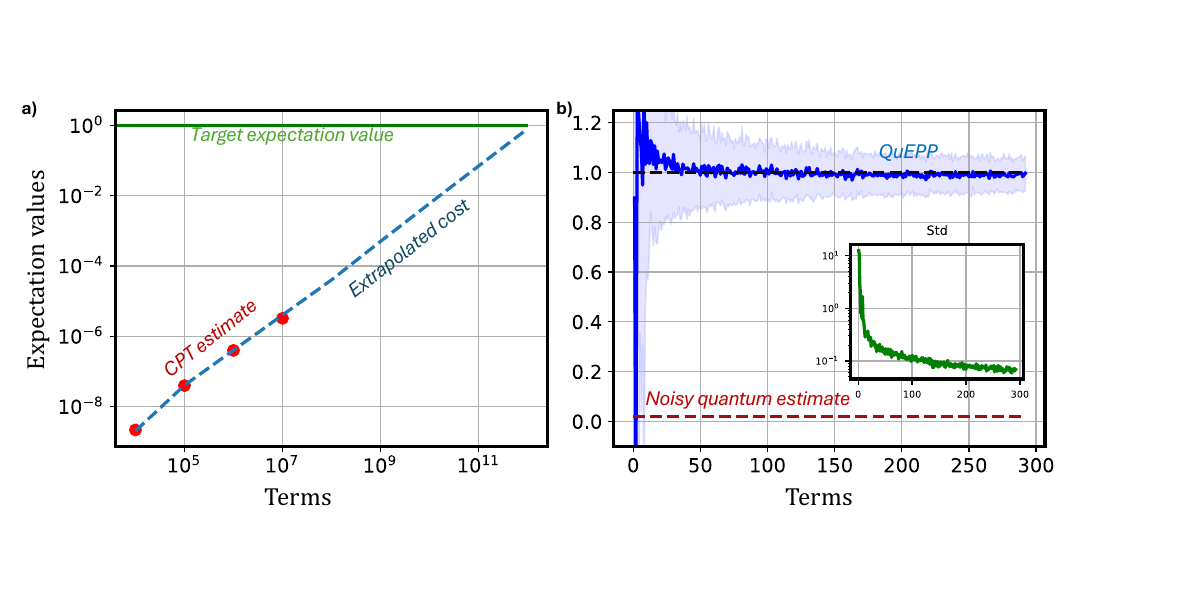}
\caption{\textbf{Unstructured mirror circuit}
The target circuit consists of preparing all zero state on 40 qubits and 30 layers of a forward propagating random 1D circuit  followed by the mirror inverse circuit. We are interested in measuring the expectation value of the observable $\langle Z_{1}Z_{4}Z_{12}Z_{13}Z_{15}Z_{21}Z_{26}Z_{27}Z_{28}Z_{32}Z_{33}Z_{34}Z_{37}Z_{39} \rangle$. a) Expectation value as a function of number of terms kept using CPT. We perform CPT until $10^{7}$ terms and then extrapolate to the ideal expectation values. We expect CPT will require $10^{12}$ terms to converge. b) Expectation value as a function of number of terms for QuEPP with Monte Carlo sampling method  converges to the $0.99\pm 0.07$ within 293 terms. The inset shows the decay of the standard error of the mean of the estimator as a function of number of terms kept. We also plot the unmitigated expectation value as a red dotted line. } 
\label{fig3:Random Mirror-MC}
\end{figure*}

The first experiment we consider is a pseudo-random 2D mirror circuit on 49 qubits and two-qubit depth of 32. The forward evolution consists of 16 layers of gates drawn from the single-qubit Clifford gates \{$H$\}, the two-qubit $CZ$ gate, and $RX$ rotations with a fixed (non-Clifford) angle. Specifically, the total circuit (combining forward and backward evolution) had 432 $CZ$ gates, 342 $H$ gates and 50 $RX(\frac{\pi}{5})$ rotations. After the circuit is applied, we measure $\langle Z_{0}Z_{11}Z_{18}Z_{41}Z_{48} \rangle$. Because of the mirror structure of the circuit, the ideal expectation value is known be 1, and as such this experiment serves as a classically verifiable problem. In total, the CPT ensemble consisted of 17 Clifford circuits, and these circuits as well as the target non-Clifford mirror circuit were executed with 100 Pauli twirled instances and 2000 shots per instance on the quantum computer.

In Fig.~\ref{fig3:Random Mirror} we show the results applying QuEPP to this circuit. The bulk of the performance improvement in Fig \ref{fig3:Random Mirror} from QuEPP is due to the initial error mitigation by rescaling, similar to Clifford data regression \cite{Czarnik2021}, and in some sense QuEPP determines the correct ensemble to use when determining the rescaling parameter. In the supplementary material we discuss ways to bound the remaining bias after truncating the QuEPP protocol at finite order, but here we emphasize that QuEPP gives a clear and rigorous procedure to systematically reduce bias by increasing resources (ensemble size), which improves confidence that its result is correct. This is clearly shown in Fig.~\ref{fig3:Random Mirror}b), which has a nearly monotonic decrease in remaining bias as the number of terms in the ensemble increases.
 
\subsubsection{Hard-to-simulate regime}

The second experiment we ran is another pseudo-random unstructured mirror circuit on a 1D chain with 32 qubits and two-qubit depth of 80. This is a significantly deeper circuit than our previous 49 qubit example. The forward evolution consists of 40 layers of gates drawn from the single-qubit Clifford gates \{$H$, $S$, $S_{dag}$\}, the two-qubit $CZ$ gate, and $RX$ rotations with a fixed (non-Clifford) angle. Specifically, the total circuit (combining forward and backward evolution) had 610 $CZ$ gates, with 400 $RX(\frac{\pi}{5})$ rotations. Here we measure a high weight observable $\langle Z_{1}Z_{4}Z_{12}Z_{13}Z_{15}Z_{21}Z_{26}Z_{27}Z_{28}Z_{32}Z_{33}Z_{34}Z_{37}Z_{39} \rangle$. There are a few things that make this circuit harder for Pauli-propagation methods and also for other error mitigation methods. The depth and number of non-Clifford rotations makes it challenging for Pauli-propagation as the number of terms in the series sum grows significantly. We used coefficient truncation based Pauli-propagation code (where we merge different paths that result in the same Pauli --- discussed further later) to classically simulate this circuit and estimated (Fig.~\ref{fig3:Random Mirror-MC} (a)) that one needs over a trillion Pauli paths to estimate the target expectation value accurately. This circuit is also unstructured and has many different unique two-qubit layers, and as such is challenging for noise learning based error mitigation methods. 

To calculate the expectation value with QuEPP, we first sampled 293 Clifford paths with Monte Carlo sampling, and then ran each path with 100 Pauli twirling instances and 200 shots each. In Fig \ref{fig3:Random Mirror-MC} (b) we plot the expectation value as a function of the number of circuits. As can be seen, the QuEPP mitigated result quickly converges to a high level of accuracy. While this result may seem impressive, we caution that in this regime of operation QuEPP, like most other QEM protocols, is a heuristic, but with a large amount of quantum execution data (the CPT ensemble) that can be used to build confidence in the result. 

\subsection{Hamiltonian Evolution}
\newcommand{\CZij}[2]{\mathrm{CZ}_{#1,#2}}
\newcommand{\Rx}[2]{R_x^{(#1)}\!\left(#2\right)}

\newcommand{\CZodd}[1]{\prod_{\substack{i=1\\ i\ \text{odd}}}^{#1-1} \CZij{i}{\,i+1}}
\newcommand{\CZeven}[1]{\prod_{\substack{i=1\\ i\ \text{even}}}^{#1-1} \CZij{i}{\,i+1}}

\newcommand{\Cell}[3]{%
  \CZodd{#1}\;
  \Big(\!\bigotimes_{j \in \mathcal{A}_{#3}} \Rx{j}{\tfrac{\pi}{2}}\!\Big)\; \\
  \CZeven{#1}\;
  \Big(\!\bigotimes_{j \in \mathcal{B}_{#3}} \Rx{j}{\theta_{j,#3}}\!\Big)
}

\newcommand{\HAll}{H^{\otimes n}}
\newcommand{\CZEven}{\prod_{\substack{i=0 \\ i\text{ even}}}^{n-2} CZ_{i,i+1}}
\newcommand{\CZOdd}{\prod_{\substack{i=1 \\ i\text{ odd}}}^{n-2} CZ_{i,i+1}}
\newcommand{\SXEven}{\prod_{\substack{i=1 \\ i\text{ odd}}}^{n-1} S_X^{(i)}}
\newcommand{\SXOdd}{\prod_{\substack{i=3 \\ i\text{ odd}}}^{n-1} S_X^{(i)}}
\newcommand{\RXAll}{\prod_{i=0}^{n-1} R_X^{(i)}(\theta)}

\begin{figure*}[t!]
\setlength{\tabcolsep}{10pt}
\centering
\includegraphics[width=\textwidth]{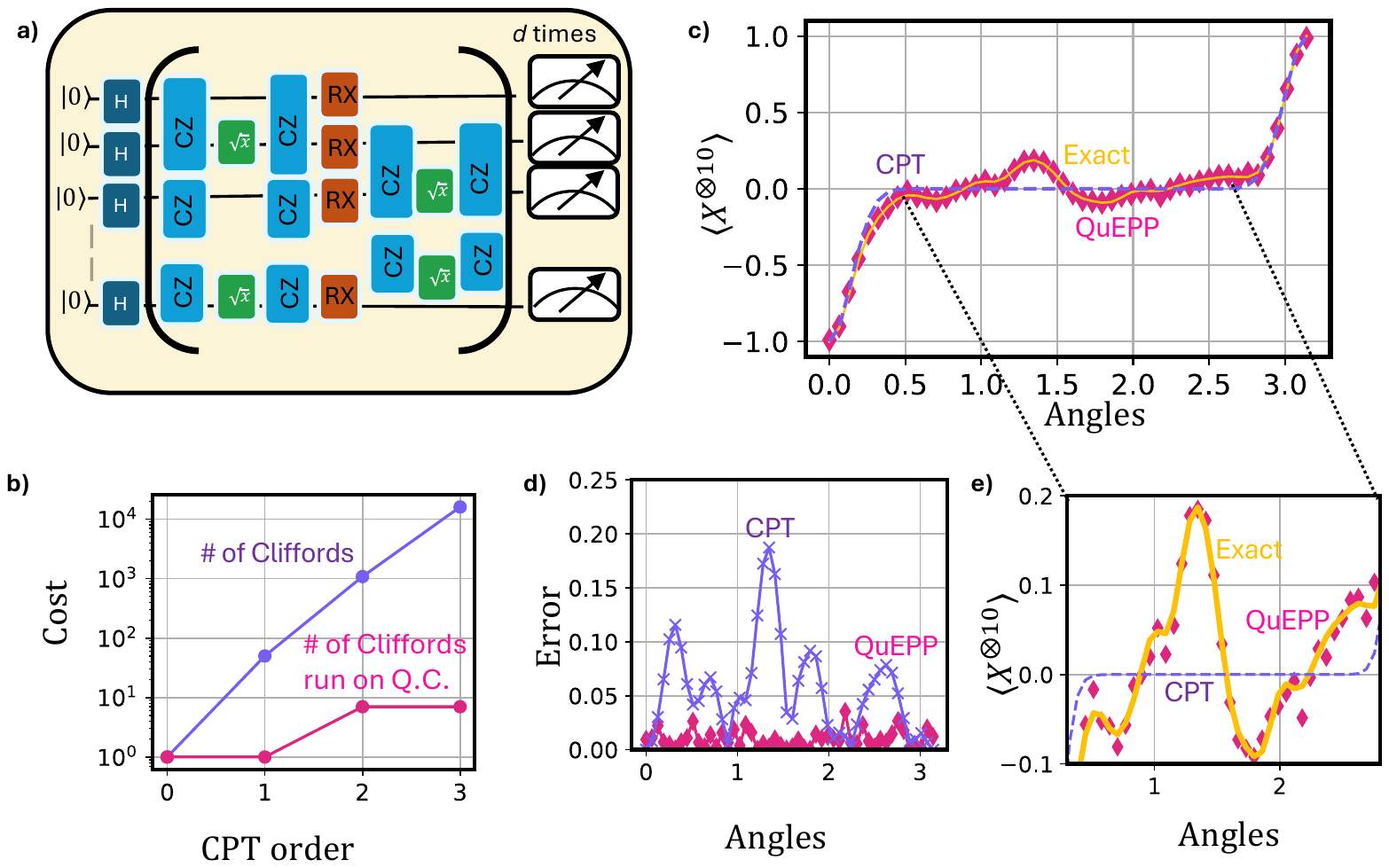}
\caption{\textbf{Trotterized Hamiltonian Time Evolution
} a) One layer of the 10 qubit Trotterized circuit is shown. The circuit is parameterized by RX rotations between the entangling layers. The circuit being run repeats this layer 10 times. All qubits are initialized in $|+\rangle$ and finally we measure $\langle X^{\otimes10} \rangle$. b) Number of Clifford paths as a function of CPT order (plotted in purple) and number of these circuits being run on a quantum computer (in Pink). c) $\langle X^{\otimes10}$ plotted as a function of the RX rotation angle for ideal (simulated using state vector simulator - in green), CPT (truncated at order 3 - in dotted Purple) and QuEPP (with CPT truncated at order 3 - in pink). d) Estimated difference between exact expectation value and the ones predicted by CPT/QuEPP plotted as a function of the rotation angle. e) Zoomed in view of (c) focusing on the region where CPT alone struggles to predict the expectation value but QuEPP can recover the signal well. }
\label{fig4:Hamiltonian}
\end{figure*}

For our second example, we execute a circuit which has the form of Trotterized time evolution as used in quantum simulation. Here, we prepare 10 qubits in the $\ket{+}^{\otimes 10}$ state, and then evolve under the unitary \[
\begin{aligned}
U = {} & \CZOdd \, \SXOdd \, \CZOdd \, \RXAll \\
       & \times \CZEven \, \SXEven \, \CZEven \, \HAll
\end{aligned}
.\] Finally, we measure a full-weight observable, $\langle X \rangle ^{\otimes 10}$. The time-evolution unitary is parameterized by $RX(\theta)$ rotations applied to all qubits between even and odd entangling layers, and we sweep this parameter $\theta \in \{0, \pi\}$ at 50 equally spaced discrete points. Using the CPT ensemble up to order $K_T = 3$, we estimate the expectation value with CPT and QuEPP, and the results are seen in Fig.~\ref{fig4:Hamiltonian}. Each circuit was run with 100 Pauli twirled instances with 200 shots each. We find that at this low order, CPT is only able to reproduce the ideal expectation for values of $\theta$ to correspond to near Clifford circuits, whereas QuEPP accurately estimates the ideal signal for all values of $\theta$.

\section{Conclusion}
\label{sec:conc}

In this work, we have proposed and demonstrated QuEPP, a quantum error mitigation protocol that can be applied to any quantum circuit, with no restriction on circuit structure. Unlike many state-of-the-art QEM protocols, QuEPP does not require any classical pre-processing of quantum circuit data (such as noise-model learning), is guaranteed to converge to the ideal value (under weak assumptions about circuit noise), and has a built-in method for bounding residual unmitigated bias error. As error mitigation is performed after all quantum circuit executions, these executions can be interleaved to minimize the impact of noise instability and drift. 

QuEPP is the simplest-to-implement variant of the Boosted Error Mitigation framework introduced in the supplementary material, which can systematically improve the accuracy of almost any error mitigation procedure. Natural directions for future work include Boosted Error Mitigation protocols for circuit ensemble that are not Clifford, but for which each circuit is still classically efficient to simulate (such as with tensor networks), and QuEPP with a QEM protocol more complicated than rescaling, such as probabilistic error cancellation \cite{Berg2023}. Such an approach would be beneficial if it resulted in a smaller ensemble, more consistent noise across ensemble circuits, or more accurate mitigated results.

Another direction for extending QuEPP is the choice of the ensemble subset beyond weighting by coefficient magnitude (series order). For example, an approach based on Pauli-path light-cone techniques \cite{Eddins24} that selects those ensemble circuits that are the most corrupted by noise in the unmitigated circuit execution. Similarly, CPT with order (coefficient) based truncation can be formulated with tree search algorithms with two primary variants: depth first search (DFS) and breadth first search (BFS) \cite{rudolph25}. Our implementation of QuEPP can only be run with DFS because unlike standard CPT in BFS mode, QuEPP cannot merge paths that result in the same Pauli and must treat them independently, as how circuit noise impacts these paths may vary. We do not believe this is a significant disadvantage because, as the authors of Ref.~\cite{rudolph25} correctly point out, BFS approaches will run out of memory very quickly. QuEPP can be extended to work in BFS mode by sampling a representative Clifford circuit from the merged paths and using its scale factor for all paths in the merge. This comes at the cost of adding the additional assumption that circuit noise across the merged paths is sufficiently uniform such that sampling only one of them does not introduce significant bias error. Monte-Carlo (MC) based DFS methods are also a promising path forward for QuEPP. This is because for circuits with many non-Clifford gates, as seen in Fig.~\ref{fig4:Hamiltonian}, the contribution from each classically simulable path becomes very small, and as such the accuracy (and unbiased-ness) of the rescaling parameters, $\eta_{i,k}$, becomes a significant concern. In MC QuEPP, we compute the rescaling parameter distribution based on probabilistic sampling from all Clifford paths, and as such it includes impact from both lower-order and higher-order paths according to their contribution to the final expectation value. This is unlike BFS methods where we discard paths based on either their coefficient or their order.     

QuEPP is a flagship example of the quantum centric supercomputing framework, as it leverages high-performance classical circuit simulation to boost the quality of an error mitigated quantum circuit execution. For future quantum computers with more consistent noise properties, it will naturally fit into a distributed quantum computing framework, as ensemble circuits can be parallelized across distributed quantum processing units to reduce individual load.
It can be used simultaneously with other low-overhead QEM protocols like symmetry-based post-selection \cite{Robledo-Moreno25}, which would both further enhance performance and lower the overall computational overhead. 

Our demonstrations in this work have focused on benchmarking the performance of QuEPP with verifiable random circuit constructions. One useful application domain for QuEPP will be quantum simulation tasks with circuit structure too complex for learning-based mitigation strategies. Another domain include tasks that execute multiple variations of a fixed ``template'' circuit, such as the variational quantum algorithms used in quantum chemistry and quantum machine learning \cite{Cerezo21}. If the variational parameters are encoded in Pauli-rotation gates, then QuEPP can be applied to all the variational circuits with a single CPT ensemble set by the circuit template. This drastically reduces the mitigation overhead for common algorithms such as the variational quantum eigensolver \cite{Peruzzo2014} or quantum kernel methods \cite{Havlek2019,Schuld19}.
With its uniquely broad scope of applicability and ease of implementation, we expect QuEPP to be a central protocol in the quest for quantum advantage in the near-term and beyond.

\bibliography{BEM_bib}

\begin{thebibliography}{32}%
\makeatletter
\providecommand \@ifxundefined [1]{%
 \@ifx{#1\undefined}
}%
\providecommand \@ifnum [1]{%
 \ifnum #1\expandafter \@firstoftwo
 \else \expandafter \@secondoftwo
 \fi
}%
\providecommand \@ifx [1]{%
 \ifx #1\expandafter \@firstoftwo
 \else \expandafter \@secondoftwo
 \fi
}%
\providecommand \natexlab [1]{#1}%
\providecommand \enquote  [1]{``#1''}%
\providecommand \bibnamefont  [1]{#1}%
\providecommand \bibfnamefont [1]{#1}%
\providecommand \citenamefont [1]{#1}%
\providecommand \href@noop [0]{\@secondoftwo}%
\providecommand \href [0]{\begingroup \@sanitize@url \@href}%
\providecommand \@href[1]{\@@startlink{#1}\@@href}%
\providecommand \@@href[1]{\endgroup#1\@@endlink}%
\providecommand \@sanitize@url [0]{\catcode `\\12\catcode `\$12\catcode `\&12\catcode `\#12\catcode `\^12\catcode `\_12\catcode `\%12\relax}%
\providecommand \@@startlink[1]{}%
\providecommand \@@endlink[0]{}%
\providecommand \url  [0]{\begingroup\@sanitize@url \@url }%
\providecommand \@url [1]{\endgroup\@href {#1}{\urlprefix }}%
\providecommand \urlprefix  [0]{URL }%
\providecommand \Eprint [0]{\href }%
\providecommand \doibase [0]{https://doi.org/}%
\providecommand \selectlanguage [0]{\@gobble}%
\providecommand \bibinfo  [0]{\@secondoftwo}%
\providecommand \bibfield  [0]{\@secondoftwo}%
\providecommand \translation [1]{[#1]}%
\providecommand \BibitemOpen [0]{}%
\providecommand \bibitemStop [0]{}%
\providecommand \bibitemNoStop [0]{.\EOS\space}%
\providecommand \EOS [0]{\spacefactor3000\relax}%
\providecommand \BibitemShut  [1]{\csname bibitem#1\endcsname}%
\let\auto@bib@innerbib\@empty
\bibitem [{\citenamefont {Kim}\ \emph {et~al.}(2023)\citenamefont {Kim}, \citenamefont {Eddins}, \citenamefont {Anand}, \citenamefont {Wei}, \citenamefont {van~den Berg}, \citenamefont {Rosenblatt}, \citenamefont {Nayfeh}, \citenamefont {Wu}, \citenamefont {Zaletel}, \citenamefont {Temme},\ and\ \citenamefont {Kandala}}]{utility}%
  \BibitemOpen
  \bibfield  {author} {\bibinfo {author} {\bibfnamefont {Y.}~\bibnamefont {Kim}}, \bibinfo {author} {\bibfnamefont {A.}~\bibnamefont {Eddins}}, \bibinfo {author} {\bibfnamefont {S.}~\bibnamefont {Anand}}, \bibinfo {author} {\bibfnamefont {K.~X.}\ \bibnamefont {Wei}}, \bibinfo {author} {\bibfnamefont {E.}~\bibnamefont {van~den Berg}}, \bibinfo {author} {\bibfnamefont {S.}~\bibnamefont {Rosenblatt}}, \bibinfo {author} {\bibfnamefont {H.}~\bibnamefont {Nayfeh}}, \bibinfo {author} {\bibfnamefont {Y.}~\bibnamefont {Wu}}, \bibinfo {author} {\bibfnamefont {M.}~\bibnamefont {Zaletel}}, \bibinfo {author} {\bibfnamefont {K.}~\bibnamefont {Temme}},\ and\ \bibinfo {author} {\bibfnamefont {A.}~\bibnamefont {Kandala}},\ }\bibfield  {title} {\bibinfo {title} {Evidence for the utility of quantum computing before fault tolerance},\ }\href@noop {} {\bibfield  {journal} {\bibinfo  {journal} {Nature}\ }\textbf {\bibinfo {volume} {618}},\ \bibinfo {pages} {500} (\bibinfo {year} {2023})}\BibitemShut {NoStop}%
\bibitem [{\citenamefont {Cai}\ \emph {et~al.}(2023)\citenamefont {Cai}, \citenamefont {Babbush}, \citenamefont {Benjamin}, \citenamefont {Endo}, \citenamefont {Huggins}, \citenamefont {Li}, \citenamefont {McClean},\ and\ \citenamefont {O'Brien}}]{Cai23}%
  \BibitemOpen
  \bibfield  {author} {\bibinfo {author} {\bibfnamefont {Z.}~\bibnamefont {Cai}}, \bibinfo {author} {\bibfnamefont {R.}~\bibnamefont {Babbush}}, \bibinfo {author} {\bibfnamefont {S.~C.}\ \bibnamefont {Benjamin}}, \bibinfo {author} {\bibfnamefont {S.}~\bibnamefont {Endo}}, \bibinfo {author} {\bibfnamefont {W.~J.}\ \bibnamefont {Huggins}}, \bibinfo {author} {\bibfnamefont {Y.}~\bibnamefont {Li}}, \bibinfo {author} {\bibfnamefont {J.~R.}\ \bibnamefont {McClean}},\ and\ \bibinfo {author} {\bibfnamefont {T.~E.}\ \bibnamefont {O'Brien}},\ }\bibfield  {title} {\bibinfo {title} {Quantum error mitigation},\ }\href {https://doi.org/10.1103/RevModPhys.95.045005} {\bibfield  {journal} {\bibinfo  {journal} {Rev. Mod. Phys.}\ }\textbf {\bibinfo {volume} {95}},\ \bibinfo {pages} {045005} (\bibinfo {year} {2023})}\BibitemShut {NoStop}%
\bibitem [{\citenamefont {Begušić}\ \emph {et~al.}(2023)\citenamefont {Begušić}, \citenamefont {Hejazi},\ and\ \citenamefont {Chan}}]{Begusic23}%
  \BibitemOpen
  \bibfield  {author} {\bibinfo {author} {\bibfnamefont {T.}~\bibnamefont {Begušić}}, \bibinfo {author} {\bibfnamefont {K.}~\bibnamefont {Hejazi}},\ and\ \bibinfo {author} {\bibfnamefont {G.~K.-L.}\ \bibnamefont {Chan}},\ }\href@noop {} {\bibinfo {title} {Simulating quantum circuit expectation values by clifford perturbation theory}} (\bibinfo {year} {2023}),\ \Eprint {https://arxiv.org/abs/arXiv:2306.04797} {arXiv:2306.04797} \BibitemShut {NoStop}%
\bibitem [{\citenamefont {Rudolph}\ \emph {et~al.}(2025)\citenamefont {Rudolph}, \citenamefont {Jones}, \citenamefont {Teng}, \citenamefont {Angrisani},\ and\ \citenamefont {Holmes}}]{rudolph25}%
  \BibitemOpen
  \bibfield  {author} {\bibinfo {author} {\bibfnamefont {M.~S.}\ \bibnamefont {Rudolph}}, \bibinfo {author} {\bibfnamefont {T.}~\bibnamefont {Jones}}, \bibinfo {author} {\bibfnamefont {Y.}~\bibnamefont {Teng}}, \bibinfo {author} {\bibfnamefont {A.}~\bibnamefont {Angrisani}},\ and\ \bibinfo {author} {\bibfnamefont {Z.}~\bibnamefont {Holmes}},\ }\bibfield  {title} {\bibinfo {title} {Pauli propagation: A computational framework for simulating quantum systems},\ }\href {https://arxiv.org/abs/2501.13101} {\bibfield  {journal} {\bibinfo  {journal} {arXiv preprint arXiv:2501.13101}\ } (\bibinfo {year} {2025})}\BibitemShut {NoStop}%
\bibitem [{\citenamefont {Liu}\ \emph {et~al.}(2024)\citenamefont {Liu}, \citenamefont {Xiao},\ and\ \citenamefont {Cai}}]{Liu24}%
  \BibitemOpen
  \bibfield  {author} {\bibinfo {author} {\bibfnamefont {Z.}~\bibnamefont {Liu}}, \bibinfo {author} {\bibfnamefont {Y.}~\bibnamefont {Xiao}},\ and\ \bibinfo {author} {\bibfnamefont {Z.}~\bibnamefont {Cai}},\ }\href@noop {} {\bibinfo {title} {Non-markovian noise suppression simplified through channel representation}} (\bibinfo {year} {2024}),\ \Eprint {https://arxiv.org/abs/arXiv:2412.11220} {arXiv:2412.11220} \BibitemShut {NoStop}%
\bibitem [{\citenamefont {Wang}\ and\ \citenamefont {Li}(2025)}]{Wang25}%
  \BibitemOpen
  \bibfield  {author} {\bibinfo {author} {\bibfnamefont {K.}~\bibnamefont {Wang}}\ and\ \bibinfo {author} {\bibfnamefont {X.}~\bibnamefont {Li}},\ }\href@noop {} {\bibinfo {title} {Non-markovian noise mitigation: Practical implementation, error analysis, and the role of environment spectral properties}} (\bibinfo {year} {2025}),\ \Eprint {https://arxiv.org/abs/arXiv:2501.05019} {arXiv:2501.05019} \BibitemShut {NoStop}%
\bibitem [{\citenamefont {van~den Berg}\ \emph {et~al.}(2023)\citenamefont {van~den Berg}, \citenamefont {Minev}, \citenamefont {Kandala},\ and\ \citenamefont {Temme}}]{Berg2023}%
  \BibitemOpen
  \bibfield  {author} {\bibinfo {author} {\bibfnamefont {E.}~\bibnamefont {van~den Berg}}, \bibinfo {author} {\bibfnamefont {Z.~K.}\ \bibnamefont {Minev}}, \bibinfo {author} {\bibfnamefont {A.}~\bibnamefont {Kandala}},\ and\ \bibinfo {author} {\bibfnamefont {K.}~\bibnamefont {Temme}},\ }\bibfield  {title} {\bibinfo {title} {Probabilistic error cancellation with sparse pauli--lindblad models on noisy quantum processors},\ }\href {https://doi.org/10.1038/s41567-023-02042-2} {\bibfield  {journal} {\bibinfo  {journal} {Nature Physics}\ }\textbf {\bibinfo {volume} {19}},\ \bibinfo {pages} {1116} (\bibinfo {year} {2023})}\BibitemShut {NoStop}%
\bibitem [{\citenamefont {Filippov}\ \emph {et~al.}(2023)\citenamefont {Filippov}, \citenamefont {Leahy}, \citenamefont {Rossi},\ and\ \citenamefont {García-Pérez}}]{Filippov23}%
  \BibitemOpen
  \bibfield  {author} {\bibinfo {author} {\bibfnamefont {S.}~\bibnamefont {Filippov}}, \bibinfo {author} {\bibfnamefont {M.}~\bibnamefont {Leahy}}, \bibinfo {author} {\bibfnamefont {M.~A.~C.}\ \bibnamefont {Rossi}},\ and\ \bibinfo {author} {\bibfnamefont {G.}~\bibnamefont {García-Pérez}},\ }\href@noop {} {\bibinfo {title} {Scalable tensor-network error mitigation for near-term quantum computing}} (\bibinfo {year} {2023}),\ \Eprint {https://arxiv.org/abs/arXiv:2307.11740} {arXiv:2307.11740} \BibitemShut {NoStop}%
\bibitem [{\citenamefont {Czarnik}\ \emph {et~al.}(2021)\citenamefont {Czarnik}, \citenamefont {Arrasmith}, \citenamefont {Coles},\ and\ \citenamefont {Cincio}}]{Czarnik2021}%
  \BibitemOpen
  \bibfield  {author} {\bibinfo {author} {\bibfnamefont {P.}~\bibnamefont {Czarnik}}, \bibinfo {author} {\bibfnamefont {A.}~\bibnamefont {Arrasmith}}, \bibinfo {author} {\bibfnamefont {P.~J.}\ \bibnamefont {Coles}},\ and\ \bibinfo {author} {\bibfnamefont {L.}~\bibnamefont {Cincio}},\ }\bibfield  {title} {\bibinfo {title} {Error mitigation with {C}lifford quantum-circuit data},\ }\href {https://doi.org/10.22331/q-2021-11-26-592} {\bibfield  {journal} {\bibinfo  {journal} {{Quantum}}\ }\textbf {\bibinfo {volume} {5}},\ \bibinfo {pages} {592} (\bibinfo {year} {2021})}\BibitemShut {NoStop}%
\bibitem [{\citenamefont {Farrell}\ \emph {et~al.}(2024)\citenamefont {Farrell}, \citenamefont {Illa}, \citenamefont {Ciavarella},\ and\ \citenamefont {Savage}}]{Farrell24_1}%
  \BibitemOpen
  \bibfield  {author} {\bibinfo {author} {\bibfnamefont {R.~C.}\ \bibnamefont {Farrell}}, \bibinfo {author} {\bibfnamefont {M.}~\bibnamefont {Illa}}, \bibinfo {author} {\bibfnamefont {A.~N.}\ \bibnamefont {Ciavarella}},\ and\ \bibinfo {author} {\bibfnamefont {M.~J.}\ \bibnamefont {Savage}},\ }\bibfield  {title} {\bibinfo {title} {Scalable circuits for preparing ground states on digital quantum computers: The schwinger model vacuum on 100 qubits},\ }\href {https://doi.org/10.1103/PRXQuantum.5.020315} {\bibfield  {journal} {\bibinfo  {journal} {PRX Quantum}\ }\textbf {\bibinfo {volume} {5}},\ \bibinfo {pages} {020315} (\bibinfo {year} {2024})}\BibitemShut {NoStop}%
\bibitem [{\citenamefont {Li}\ and\ \citenamefont {Benjamin}(2017)}]{Li17}%
  \BibitemOpen
  \bibfield  {author} {\bibinfo {author} {\bibfnamefont {Y.}~\bibnamefont {Li}}\ and\ \bibinfo {author} {\bibfnamefont {S.~C.}\ \bibnamefont {Benjamin}},\ }\bibfield  {title} {\bibinfo {title} {Efficient variational quantum simulator incorporating active error minimization},\ }\href {https://doi.org/10.1103/PhysRevX.7.021050} {\bibfield  {journal} {\bibinfo  {journal} {Phys. Rev. X}\ }\textbf {\bibinfo {volume} {7}},\ \bibinfo {pages} {021050} (\bibinfo {year} {2017})}\BibitemShut {NoStop}%
\bibitem [{\citenamefont {Endo}\ \emph {et~al.}(2018)\citenamefont {Endo}, \citenamefont {Benjamin},\ and\ \citenamefont {Li}}]{Endo18}%
  \BibitemOpen
  \bibfield  {author} {\bibinfo {author} {\bibfnamefont {S.}~\bibnamefont {Endo}}, \bibinfo {author} {\bibfnamefont {S.~C.}\ \bibnamefont {Benjamin}},\ and\ \bibinfo {author} {\bibfnamefont {Y.}~\bibnamefont {Li}},\ }\bibfield  {title} {\bibinfo {title} {Practical quantum error mitigation for near-future applications},\ }\href {https://doi.org/10.1103/PhysRevX.8.031027} {\bibfield  {journal} {\bibinfo  {journal} {Phys. Rev. X}\ }\textbf {\bibinfo {volume} {8}},\ \bibinfo {pages} {031027} (\bibinfo {year} {2018})}\BibitemShut {NoStop}%
\bibitem [{\citenamefont {Govia}\ \emph {et~al.}(2025)\citenamefont {Govia}, \citenamefont {Majumder}, \citenamefont {Barron}, \citenamefont {Mitchell}, \citenamefont {Seif}, \citenamefont {Kim}, \citenamefont {Wood}, \citenamefont {Pritchett}, \citenamefont {Merkel},\ and\ \citenamefont {McKay}}]{Govia25}%
  \BibitemOpen
  \bibfield  {author} {\bibinfo {author} {\bibfnamefont {L.}~\bibnamefont {Govia}}, \bibinfo {author} {\bibfnamefont {S.}~\bibnamefont {Majumder}}, \bibinfo {author} {\bibfnamefont {S.}~\bibnamefont {Barron}}, \bibinfo {author} {\bibfnamefont {B.}~\bibnamefont {Mitchell}}, \bibinfo {author} {\bibfnamefont {A.}~\bibnamefont {Seif}}, \bibinfo {author} {\bibfnamefont {Y.}~\bibnamefont {Kim}}, \bibinfo {author} {\bibfnamefont {C.}~\bibnamefont {Wood}}, \bibinfo {author} {\bibfnamefont {E.}~\bibnamefont {Pritchett}}, \bibinfo {author} {\bibfnamefont {S.}~\bibnamefont {Merkel}},\ and\ \bibinfo {author} {\bibfnamefont {D.}~\bibnamefont {McKay}},\ }\bibfield  {title} {\bibinfo {title} {Bounding the systematic error in quantum error mitigation due to model violation},\ }\href {https://doi.org/10.1103/PRXQuantum.6.010354} {\bibfield  {journal} {\bibinfo  {journal} {PRX Quantum}\ }\textbf {\bibinfo {volume} {6}},\ \bibinfo {pages} {010354} (\bibinfo {year} {2025})}\BibitemShut {NoStop}%
\bibitem [{\citenamefont {Filippov}\ \emph {et~al.}(2024)\citenamefont {Filippov}, \citenamefont {Maniscalco},\ and\ \citenamefont {García-Pérez}}]{Filippov24}%
  \BibitemOpen
  \bibfield  {author} {\bibinfo {author} {\bibfnamefont {S.~N.}\ \bibnamefont {Filippov}}, \bibinfo {author} {\bibfnamefont {S.}~\bibnamefont {Maniscalco}},\ and\ \bibinfo {author} {\bibfnamefont {G.}~\bibnamefont {García-Pérez}},\ }\href@noop {} {\bibinfo {title} {Scalability of quantum error mitigation techniques: from utility to advantage}} (\bibinfo {year} {2024}),\ \Eprint {https://arxiv.org/abs/arXiv:2403.13542} {arXiv:2403.13542} \BibitemShut {NoStop}%
\bibitem [{\citenamefont {Kim}\ \emph {et~al.}(2024)\citenamefont {Kim}, \citenamefont {Govia}, \citenamefont {Dane}, \citenamefont {van~den Berg}, \citenamefont {Zajac}, \citenamefont {Mitchell}, \citenamefont {Liu}, \citenamefont {Balakrishnan}, \citenamefont {Keefe}, \citenamefont {Stabile}, \citenamefont {Pritchett}, \citenamefont {Stehlik},\ and\ \citenamefont {Kandala}}]{kim2024}%
  \BibitemOpen
  \bibfield  {author} {\bibinfo {author} {\bibfnamefont {Y.}~\bibnamefont {Kim}}, \bibinfo {author} {\bibfnamefont {L.~C.~G.}\ \bibnamefont {Govia}}, \bibinfo {author} {\bibfnamefont {A.}~\bibnamefont {Dane}}, \bibinfo {author} {\bibfnamefont {E.}~\bibnamefont {van~den Berg}}, \bibinfo {author} {\bibfnamefont {D.~M.}\ \bibnamefont {Zajac}}, \bibinfo {author} {\bibfnamefont {B.}~\bibnamefont {Mitchell}}, \bibinfo {author} {\bibfnamefont {Y.}~\bibnamefont {Liu}}, \bibinfo {author} {\bibfnamefont {K.}~\bibnamefont {Balakrishnan}}, \bibinfo {author} {\bibfnamefont {G.}~\bibnamefont {Keefe}}, \bibinfo {author} {\bibfnamefont {A.}~\bibnamefont {Stabile}}, \bibinfo {author} {\bibfnamefont {E.}~\bibnamefont {Pritchett}}, \bibinfo {author} {\bibfnamefont {J.}~\bibnamefont {Stehlik}},\ and\ \bibinfo {author} {\bibfnamefont {A.}~\bibnamefont {Kandala}},\ }\href {https://arxiv.org/abs/2407.02467} {\bibinfo {title} {Error mitigation with stabilized noise in superconducting quantum processors}} (\bibinfo {year} {2024}),\
  \Eprint {https://arxiv.org/abs/2407.02467} {arXiv:2407.02467 [quant-ph]} \BibitemShut {NoStop}%
\bibitem [{\citenamefont {Merkel}\ \emph {et~al.}(2025)\citenamefont {Merkel}, \citenamefont {Proctor}, \citenamefont {Ferracin}, \citenamefont {Hines}, \citenamefont {Barron}, \citenamefont {Govia},\ and\ \citenamefont {McKay}}]{Merkel25}%
  \BibitemOpen
  \bibfield  {author} {\bibinfo {author} {\bibfnamefont {S.}~\bibnamefont {Merkel}}, \bibinfo {author} {\bibfnamefont {T.}~\bibnamefont {Proctor}}, \bibinfo {author} {\bibfnamefont {S.}~\bibnamefont {Ferracin}}, \bibinfo {author} {\bibfnamefont {J.}~\bibnamefont {Hines}}, \bibinfo {author} {\bibfnamefont {S.}~\bibnamefont {Barron}}, \bibinfo {author} {\bibfnamefont {L.~C.~G.}\ \bibnamefont {Govia}},\ and\ \bibinfo {author} {\bibfnamefont {D.}~\bibnamefont {McKay}},\ }\href {https://arxiv.org/abs/2503.05943} {\bibinfo {title} {When clifford benchmarks are sufficient; estimating application performance with scalable proxy circuits}} (\bibinfo {year} {2025}),\ \Eprint {https://arxiv.org/abs/2503.05943} {arXiv:2503.05943 [quant-ph]} \BibitemShut {NoStop}%
\bibitem [{\citenamefont {Regula}\ \emph {et~al.}(2021)\citenamefont {Regula}, \citenamefont {Takagi},\ and\ \citenamefont {Gu}}]{Regula2021}%
  \BibitemOpen
  \bibfield  {author} {\bibinfo {author} {\bibfnamefont {B.}~\bibnamefont {Regula}}, \bibinfo {author} {\bibfnamefont {R.}~\bibnamefont {Takagi}},\ and\ \bibinfo {author} {\bibfnamefont {M.}~\bibnamefont {Gu}},\ }\bibfield  {title} {\bibinfo {title} {Operational applications of the diamond norm and related measures in quantifying the non-physicality of quantum maps},\ }\href {https://doi.org/10.22331/q-2021-08-09-522} {\bibfield  {journal} {\bibinfo  {journal} {{Quantum}}\ }\textbf {\bibinfo {volume} {5}},\ \bibinfo {pages} {522} (\bibinfo {year} {2021})}\BibitemShut {NoStop}%
\bibitem [{\citenamefont {Takagi}\ \emph {et~al.}(2022)\citenamefont {Takagi}, \citenamefont {Endo}, \citenamefont {Minagawa},\ and\ \citenamefont {Gu}}]{Takagi22}%
  \BibitemOpen
  \bibfield  {author} {\bibinfo {author} {\bibfnamefont {R.}~\bibnamefont {Takagi}}, \bibinfo {author} {\bibfnamefont {S.}~\bibnamefont {Endo}}, \bibinfo {author} {\bibfnamefont {S.}~\bibnamefont {Minagawa}},\ and\ \bibinfo {author} {\bibfnamefont {M.}~\bibnamefont {Gu}},\ }\bibfield  {title} {\bibinfo {title} {Fundamental limits of quantum error mitigation},\ }\href@noop {} {\bibfield  {journal} {\bibinfo  {journal} {npj Quantum Information}\ }\textbf {\bibinfo {volume} {8}},\ \bibinfo {pages} {114} (\bibinfo {year} {2022})}\BibitemShut {NoStop}%
\bibitem [{\citenamefont {Dowling}\ \emph {et~al.}(2025)\citenamefont {Dowling}, \citenamefont {Modi},\ and\ \citenamefont {White}}]{Dowling25}%
  \BibitemOpen
  \bibfield  {author} {\bibinfo {author} {\bibfnamefont {N.}~\bibnamefont {Dowling}}, \bibinfo {author} {\bibfnamefont {K.}~\bibnamefont {Modi}},\ and\ \bibinfo {author} {\bibfnamefont {G.~A.~L.}\ \bibnamefont {White}},\ }\href@noop {} {\bibinfo {title} {Bridging entanglement and magic resources through operator space}} (\bibinfo {year} {2025}),\ \Eprint {https://arxiv.org/abs/arXiv:2501.18679} {arXiv:2501.18679} \BibitemShut {NoStop}%
\bibitem [{\citenamefont {Bennett}\ \emph {et~al.}(1996)\citenamefont {Bennett}, \citenamefont {Brassard}, \citenamefont {Popescu}, \citenamefont {Schumacher}, \citenamefont {Smolin},\ and\ \citenamefont {Wootters}}]{Bennett96}%
  \BibitemOpen
  \bibfield  {author} {\bibinfo {author} {\bibfnamefont {C.~H.}\ \bibnamefont {Bennett}}, \bibinfo {author} {\bibfnamefont {G.}~\bibnamefont {Brassard}}, \bibinfo {author} {\bibfnamefont {S.}~\bibnamefont {Popescu}}, \bibinfo {author} {\bibfnamefont {B.}~\bibnamefont {Schumacher}}, \bibinfo {author} {\bibfnamefont {J.~A.}\ \bibnamefont {Smolin}},\ and\ \bibinfo {author} {\bibfnamefont {W.~K.}\ \bibnamefont {Wootters}},\ }\bibfield  {title} {\bibinfo {title} {Purification of noisy entanglement and faithful teleportation via noisy channels},\ }\href {https://doi.org/10.1103/PhysRevLett.76.722} {\bibfield  {journal} {\bibinfo  {journal} {Phys. Rev. Lett.}\ }\textbf {\bibinfo {volume} {76}},\ \bibinfo {pages} {722} (\bibinfo {year} {1996})}\BibitemShut {NoStop}%
\bibitem [{\citenamefont {Knill}(2004)}]{Knill04}%
  \BibitemOpen
  \bibfield  {author} {\bibinfo {author} {\bibfnamefont {E.}~\bibnamefont {Knill}},\ }\href@noop {} {\bibinfo {title} {Fault-tolerant postselected quantum computation: Threshold analysis}} (\bibinfo {year} {2004}),\ \Eprint {https://arxiv.org/abs/arXiv:quant-ph/0404104} {arXiv:quant-ph/0404104} \BibitemShut {NoStop}%
\bibitem [{\citenamefont {Kern}\ \emph {et~al.}(2005)\citenamefont {Kern}, \citenamefont {Alber},\ and\ \citenamefont {Shepelyansky}}]{Kern05}%
  \BibitemOpen
  \bibfield  {author} {\bibinfo {author} {\bibfnamefont {O.}~\bibnamefont {Kern}}, \bibinfo {author} {\bibfnamefont {G.}~\bibnamefont {Alber}},\ and\ \bibinfo {author} {\bibfnamefont {D.~L.}\ \bibnamefont {Shepelyansky}},\ }\bibfield  {title} {\bibinfo {title} {Quantum error correction of coherent errors by randomization},\ }\href@noop {} {\bibfield  {journal} {\bibinfo  {journal} {The European Physical Journal D - Atomic, Molecular, Optical and Plasma Physics}\ }\textbf {\bibinfo {volume} {32}},\ \bibinfo {pages} {153} (\bibinfo {year} {2005})}\BibitemShut {NoStop}%
\bibitem [{\citenamefont {Geller}\ and\ \citenamefont {Zhou}(2013)}]{Geller13}%
  \BibitemOpen
  \bibfield  {author} {\bibinfo {author} {\bibfnamefont {M.~R.}\ \bibnamefont {Geller}}\ and\ \bibinfo {author} {\bibfnamefont {Z.}~\bibnamefont {Zhou}},\ }\bibfield  {title} {\bibinfo {title} {Efficient error models for fault-tolerant architectures and the pauli twirling approximation},\ }\href {https://doi.org/10.1103/PhysRevA.88.012314} {\bibfield  {journal} {\bibinfo  {journal} {Phys. Rev. A}\ }\textbf {\bibinfo {volume} {88}},\ \bibinfo {pages} {012314} (\bibinfo {year} {2013})}\BibitemShut {NoStop}%
\bibitem [{\citenamefont {Wallman}\ and\ \citenamefont {Emerson}(2016)}]{Wallman16}%
  \BibitemOpen
  \bibfield  {author} {\bibinfo {author} {\bibfnamefont {J.~J.}\ \bibnamefont {Wallman}}\ and\ \bibinfo {author} {\bibfnamefont {J.}~\bibnamefont {Emerson}},\ }\bibfield  {title} {\bibinfo {title} {Noise tailoring for scalable quantum computation via randomized compiling},\ }\href {https://doi.org/10.1103/PhysRevA.94.052325} {\bibfield  {journal} {\bibinfo  {journal} {Phys. Rev. A}\ }\textbf {\bibinfo {volume} {94}},\ \bibinfo {pages} {052325} (\bibinfo {year} {2016})}\BibitemShut {NoStop}%
\bibitem [{\citenamefont {Eddins}\ \emph {et~al.}(2024)\citenamefont {Eddins}, \citenamefont {Tran},\ and\ \citenamefont {Rall}}]{Eddins24}%
  \BibitemOpen
  \bibfield  {author} {\bibinfo {author} {\bibfnamefont {A.}~\bibnamefont {Eddins}}, \bibinfo {author} {\bibfnamefont {M.~C.}\ \bibnamefont {Tran}},\ and\ \bibinfo {author} {\bibfnamefont {P.}~\bibnamefont {Rall}},\ }\href {https://arxiv.org/abs/2409.04401} {\bibinfo {title} {Lightcone shading for classically accelerated quantum error mitigation}} (\bibinfo {year} {2024}),\ \Eprint {https://arxiv.org/abs/2409.04401} {arXiv:2409.04401 [quant-ph]} \BibitemShut {NoStop}%
\bibitem [{\citenamefont {Robledo-Moreno}\ \emph {et~al.}(2025)\citenamefont {Robledo-Moreno}, \citenamefont {Motta}, \citenamefont {Haas}, \citenamefont {Javadi-Abhari}, \citenamefont {Jurcevic}, \citenamefont {Kirby}, \citenamefont {Martiel}, \citenamefont {Sharma}, \citenamefont {Sharma}, \citenamefont {Shirakawa}, \citenamefont {Sitdikov}, \citenamefont {Sun}, \citenamefont {Sung}, \citenamefont {Takita}, \citenamefont {Tran}, \citenamefont {Yunoki},\ and\ \citenamefont {Mezzacapo}}]{Robledo-Moreno25}%
  \BibitemOpen
  \bibfield  {author} {\bibinfo {author} {\bibfnamefont {J.}~\bibnamefont {Robledo-Moreno}}, \bibinfo {author} {\bibfnamefont {M.}~\bibnamefont {Motta}}, \bibinfo {author} {\bibfnamefont {H.}~\bibnamefont {Haas}}, \bibinfo {author} {\bibfnamefont {A.}~\bibnamefont {Javadi-Abhari}}, \bibinfo {author} {\bibfnamefont {P.}~\bibnamefont {Jurcevic}}, \bibinfo {author} {\bibfnamefont {W.}~\bibnamefont {Kirby}}, \bibinfo {author} {\bibfnamefont {S.}~\bibnamefont {Martiel}}, \bibinfo {author} {\bibfnamefont {K.}~\bibnamefont {Sharma}}, \bibinfo {author} {\bibfnamefont {S.}~\bibnamefont {Sharma}}, \bibinfo {author} {\bibfnamefont {T.}~\bibnamefont {Shirakawa}}, \bibinfo {author} {\bibfnamefont {I.}~\bibnamefont {Sitdikov}}, \bibinfo {author} {\bibfnamefont {R.-Y.}\ \bibnamefont {Sun}}, \bibinfo {author} {\bibfnamefont {K.~J.}\ \bibnamefont {Sung}}, \bibinfo {author} {\bibfnamefont {M.}~\bibnamefont {Takita}}, \bibinfo {author} {\bibfnamefont {M.~C.}\ \bibnamefont {Tran}}, \bibinfo {author} {\bibfnamefont
  {S.}~\bibnamefont {Yunoki}},\ and\ \bibinfo {author} {\bibfnamefont {A.}~\bibnamefont {Mezzacapo}},\ }\bibfield  {title} {\bibinfo {title} {Chemistry beyond the scale of exact diagonalization on a quantum-centric supercomputer},\ }\href {https://doi.org/10.1126/sciadv.adu9991} {\bibfield  {journal} {\bibinfo  {journal} {Science Advances}\ }\textbf {\bibinfo {volume} {11}},\ \bibinfo {pages} {eadu9991} (\bibinfo {year} {2025})},\ \Eprint {https://arxiv.org/abs/https://www.science.org/doi/pdf/10.1126/sciadv.adu9991} {https://www.science.org/doi/pdf/10.1126/sciadv.adu9991} \BibitemShut {NoStop}%
\bibitem [{\citenamefont {Cerezo}\ \emph {et~al.}(2021)\citenamefont {Cerezo}, \citenamefont {Arrasmith}, \citenamefont {Babbush}, \citenamefont {Benjamin}, \citenamefont {Endo}, \citenamefont {Fujii}, \citenamefont {McClean}, \citenamefont {Mitarai}, \citenamefont {Yuan}, \citenamefont {Cincio},\ and\ \citenamefont {Coles}}]{Cerezo21}%
  \BibitemOpen
  \bibfield  {author} {\bibinfo {author} {\bibfnamefont {M.}~\bibnamefont {Cerezo}}, \bibinfo {author} {\bibfnamefont {A.}~\bibnamefont {Arrasmith}}, \bibinfo {author} {\bibfnamefont {R.}~\bibnamefont {Babbush}}, \bibinfo {author} {\bibfnamefont {S.~C.}\ \bibnamefont {Benjamin}}, \bibinfo {author} {\bibfnamefont {S.}~\bibnamefont {Endo}}, \bibinfo {author} {\bibfnamefont {K.}~\bibnamefont {Fujii}}, \bibinfo {author} {\bibfnamefont {J.~R.}\ \bibnamefont {McClean}}, \bibinfo {author} {\bibfnamefont {K.}~\bibnamefont {Mitarai}}, \bibinfo {author} {\bibfnamefont {X.}~\bibnamefont {Yuan}}, \bibinfo {author} {\bibfnamefont {L.}~\bibnamefont {Cincio}},\ and\ \bibinfo {author} {\bibfnamefont {P.~J.}\ \bibnamefont {Coles}},\ }\bibfield  {title} {\bibinfo {title} {Variational quantum algorithms},\ }\href@noop {} {\bibfield  {journal} {\bibinfo  {journal} {Nature Reviews Physics}\ }\textbf {\bibinfo {volume} {3}},\ \bibinfo {pages} {625} (\bibinfo {year} {2021})}\BibitemShut {NoStop}%
\bibitem [{\citenamefont {Peruzzo}\ \emph {et~al.}(2014)\citenamefont {Peruzzo}, \citenamefont {McClean}, \citenamefont {Shadbolt}, \citenamefont {Yung}, \citenamefont {Zhou}, \citenamefont {Love}, \citenamefont {Aspuru-Guzik},\ and\ \citenamefont {O’Brien}}]{Peruzzo2014}%
  \BibitemOpen
  \bibfield  {author} {\bibinfo {author} {\bibfnamefont {A.}~\bibnamefont {Peruzzo}}, \bibinfo {author} {\bibfnamefont {J.}~\bibnamefont {McClean}}, \bibinfo {author} {\bibfnamefont {P.}~\bibnamefont {Shadbolt}}, \bibinfo {author} {\bibfnamefont {M.-H.}\ \bibnamefont {Yung}}, \bibinfo {author} {\bibfnamefont {X.-Q.}\ \bibnamefont {Zhou}}, \bibinfo {author} {\bibfnamefont {P.~J.}\ \bibnamefont {Love}}, \bibinfo {author} {\bibfnamefont {A.}~\bibnamefont {Aspuru-Guzik}},\ and\ \bibinfo {author} {\bibfnamefont {J.~L.}\ \bibnamefont {O’Brien}},\ }\bibfield  {title} {\bibinfo {title} {A variational eigenvalue solver on a photonic quantum processor},\ }\bibfield  {journal} {\bibinfo  {journal} {Nature Communications}\ }\textbf {\bibinfo {volume} {5}},\ \href {https://doi.org/10.1038/ncomms5213} {10.1038/ncomms5213} (\bibinfo {year} {2014})\BibitemShut {NoStop}%
\bibitem [{\citenamefont {Havlíček}\ \emph {et~al.}(2019)\citenamefont {Havlíček}, \citenamefont {Córcoles}, \citenamefont {Temme}, \citenamefont {Harrow}, \citenamefont {Kandala}, \citenamefont {Chow},\ and\ \citenamefont {Gambetta}}]{Havlek2019}%
  \BibitemOpen
  \bibfield  {author} {\bibinfo {author} {\bibfnamefont {V.}~\bibnamefont {Havlíček}}, \bibinfo {author} {\bibfnamefont {A.~D.}\ \bibnamefont {Córcoles}}, \bibinfo {author} {\bibfnamefont {K.}~\bibnamefont {Temme}}, \bibinfo {author} {\bibfnamefont {A.~W.}\ \bibnamefont {Harrow}}, \bibinfo {author} {\bibfnamefont {A.}~\bibnamefont {Kandala}}, \bibinfo {author} {\bibfnamefont {J.~M.}\ \bibnamefont {Chow}},\ and\ \bibinfo {author} {\bibfnamefont {J.~M.}\ \bibnamefont {Gambetta}},\ }\bibfield  {title} {\bibinfo {title} {Supervised learning with quantum-enhanced feature spaces},\ }\href {https://doi.org/10.1038/s41586-019-0980-2} {\bibfield  {journal} {\bibinfo  {journal} {Nature}\ }\textbf {\bibinfo {volume} {567}},\ \bibinfo {pages} {209–212} (\bibinfo {year} {2019})}\BibitemShut {NoStop}%
\bibitem [{\citenamefont {Schuld}\ and\ \citenamefont {Killoran}(2019)}]{Schuld19}%
  \BibitemOpen
  \bibfield  {author} {\bibinfo {author} {\bibfnamefont {M.}~\bibnamefont {Schuld}}\ and\ \bibinfo {author} {\bibfnamefont {N.}~\bibnamefont {Killoran}},\ }\bibfield  {title} {\bibinfo {title} {Quantum machine learning in feature hilbert spaces},\ }\href {https://doi.org/10.1103/PhysRevLett.122.040504} {\bibfield  {journal} {\bibinfo  {journal} {Phys. Rev. Lett.}\ }\textbf {\bibinfo {volume} {122}},\ \bibinfo {pages} {040504} (\bibinfo {year} {2019})}\BibitemShut {NoStop}%
\bibitem [{\citenamefont {van~den Berg}\ \emph {et~al.}(2022)\citenamefont {van~den Berg}, \citenamefont {Minev},\ and\ \citenamefont {Temme}}]{Berg22}%
  \BibitemOpen
  \bibfield  {author} {\bibinfo {author} {\bibfnamefont {E.}~\bibnamefont {van~den Berg}}, \bibinfo {author} {\bibfnamefont {Z.~K.}\ \bibnamefont {Minev}},\ and\ \bibinfo {author} {\bibfnamefont {K.}~\bibnamefont {Temme}},\ }\bibfield  {title} {\bibinfo {title} {Model-free readout-error mitigation for quantum expectation values},\ }\href {https://doi.org/10.1103/PhysRevA.105.032620} {\bibfield  {journal} {\bibinfo  {journal} {Phys. Rev. A}\ }\textbf {\bibinfo {volume} {105}},\ \bibinfo {pages} {032620} (\bibinfo {year} {2022})}\BibitemShut {NoStop}%
\bibitem [{\citenamefont {Lerch}\ \emph {et~al.}(2024)\citenamefont {Lerch}, \citenamefont {Puig}, \citenamefont {Rudolph}, \citenamefont {Angrisani}, \citenamefont {Jones}, \citenamefont {Cerezo}, \citenamefont {Thanasilp},\ and\ \citenamefont {Holmes}}]{Lerch24}%
  \BibitemOpen
  \bibfield  {author} {\bibinfo {author} {\bibfnamefont {S.}~\bibnamefont {Lerch}}, \bibinfo {author} {\bibfnamefont {R.}~\bibnamefont {Puig}}, \bibinfo {author} {\bibfnamefont {M.~S.}\ \bibnamefont {Rudolph}}, \bibinfo {author} {\bibfnamefont {A.}~\bibnamefont {Angrisani}}, \bibinfo {author} {\bibfnamefont {T.}~\bibnamefont {Jones}}, \bibinfo {author} {\bibfnamefont {M.}~\bibnamefont {Cerezo}}, \bibinfo {author} {\bibfnamefont {S.}~\bibnamefont {Thanasilp}},\ and\ \bibinfo {author} {\bibfnamefont {Z.}~\bibnamefont {Holmes}},\ }\href@noop {} {\bibinfo {title} {Efficient quantum-enhanced classical simulation for patches of quantum landscapes}} (\bibinfo {year} {2024}),\ \Eprint {https://arxiv.org/abs/arXiv:2411.19896} {arXiv:2411.19896} \BibitemShut {NoStop}%
\end{thebibliography}%

\clearpage{}
\section*{Supplementary Material}
\setcounter{figure}{0}
\setcounter{equation}{0}
\renewcommand{\thefigure}{S\arabic{figure}}
\renewcommand{\thesection}{S.\Roman{section}}
\renewcommand \theequation{S\arabic{equation}}
\renewcommand \thetable{S\arabic{@table}}
\renewcommand{\bibnumfmt}[1]{[S#1]}
\renewcommand{\citenumfont}[1]{S#1}


\global\long\def\Tr{\operatorname{Tr}}%
\global\long\def\isdef{\coloneqq}%
\global\long\def\ket#1{\left|#1\right\rangle }%
\global\long\def\bra#1{\left\langle #1\right|}%
\global\long\def\braket#1#2{\left\langle #1\middle|#2\right\rangle }%
\global\long\def\ketbra#1#2{\left|#1\vphantom{#2}\right\rangle \left\langle \vphantom{#1}#2\right|}%
\global\long\def\kb#1#2{\left|#1\vphantom{#2}\right\rangle \left\langle \vphantom{#1}#2\right|}%
\global\long\def\braOket#1#2#3{\left\langle #1\middle|#2\middle|#3\right\rangle }%
\global\long\def\kk{\rangle\!\rangle}%
\global\long\def\bb{\langle\!\langle}%
\global\long\def\Op{\operatorname{Op}}%

\section{Boosted Error Mitigation}
\label{sec:bem}

Having demonstrated the effectiveness of QuEPP in the main text, we now discuss how it is a specific realization of a broad framework we call Boosted Error Mitigation (BEM). The BEM framework can use any circuit ensemble description of a target circuit, not just a Pauli path description, and almost any QEM protocol to systematically reduce the bias in the initial error mitigated result. This ``boosting'' process turns the error mitigation procedure into one that has verifiable accuracy, with bounded bias that asymptotically approaches zero.

To do this, BEM requires that the circuit ensemble and the QEM procedure satisfy four properties:
\begin{enumerate}
    \item The expectation value for each circuit in the ensemble can be computed efficiently on a classical computer.
    
    \item Applying the QEM procedure to the target circuit is equivalent to applying it to each circuit in the ensemble.
    
    \item The mitigated expectation values can be computed efficiently using classical and quantum resources.
    
    \item The series coefficients in the circuit ensemble description of the target expectation value (c.f.~$g\left(\mathcal{C}\right)$ of the main text) can be ordered to reflect the magnitude of the contribution of each circuit in the ensemble to the target expectation value.
\end{enumerate}
Properties 1 and 4 depend on the choice of circuit ensemble, while properties 2 and 3 depend on the ensemble, the QEM protocol, and the noise on the quantum hardware. Properties 3 and 4 are not strictly required for BEM to work, but ensure that the computational overhead to boosting is not prohibitively large (property 3), and that output of BEM is trustworthy (property 4), in the sense that the remaining bias due to incomplete summation of the ensemble series decreases as the number of terms in the summation increases.

Variants of BEM differ in their choice of series ensemble and initial error mitigation protocol. These choices cannot be made independently, as properties 2 and 3 of the BEM requirements require the series ensemble and mitigation protocol to be compatible, in the sense that mitigation can be applied to each ensemble circuit. Focusing on Clifford circuit ensembles, we see that beyond the rescaling used in QuEPP, other compatible error mitigation protocols include Clifford Data Regression (CDR) \cite{Czarnik2021}, Operator Decoherence Renormalization (ODR) \cite{Farrell24_1}, Probabilistic Error Cancellation (PEC) \cite{Berg2023}, Probabilistic Error Amplification (PEA) \cite{utility}, and Tensor Error Mitigation (TEM) \cite{Filippov23}.

\subsection{QuEPP as a BEM protocol}

Before giving a general description of BEM we first connect back to QuEPP to explain how it fits into the BEM framework. Regarding the BEM requirements, we see that as each circuit in the CPT ensemble is a Clifford circuit, property 1 is satisfied. Given the form of the CPT coefficients, we see that a CPT ensemble satisfies property 4 as the coefficients of each order $k$ follow $g(i,k) \propto \sin^k{\theta}$ with $\abs{\sin{\theta}} < 1$, so the circuits of each order contribute less \emph{per circuit} to the overall expectation value.

The circuit-independent rescaling QEM protocol used by QuEPP satisfies property 3 as we run each Clifford circuit in the ensemble on the noisy hardware. Satisfying property 2 is a bit more nuanced. Rescaling itself trivially satisfies property 2, as rescaling the target circuit is equivalent to rescaling the ensemble sum. However, the noisy target expectation value can only be written as a sum of the noisy circuit ensemble expectation values if noise on the quantum hardware is described by a linear map that does not depend on the specific circuit. This is the common assumption behind almost all error mitigation protocols \cite{Govia25}, and is weaker even than a Markovian assumption on the noise channel \cite{Merkel25}. 

For QuEPP, we only require the even weaker assumption that replacing the non-Clifford gates with Clifford gates does not change the noise channel. While not guaranteed to be true, this is more likely to be the case when the non-Clifford gates are single qubit gates all built from the same, fixed-depth, gate decomposition. In practice, when deploying QuEPP we use Pauli twirling to force the noise channel to be a Pauli stochastic channel, which both helps satisfy property 2 and improves the convergence of QuEPP by removing the impact of coherent error.

\subsection{The Boosted Error Mitigation Protocol}

The Boosted Error Mitigation protocol consists of the following steps, which do not necessarily occur sequentially, as depending on the specific choice of series ensemble and error mitigation protocol some steps can occur simultaneously (as with QuEPP). We assume we have an initial mitigation protocol $\mathbb{M}$, and a strict subset $\mathscr{B} \subset \mathscr{E}$ of the series ensemble circuits which we will use for boosting. Note that if $\mathscr{B} = \mathscr{E}$ then BEM is equivalent to classical simulation, hence why it must be a strict subset. The BEM steps are:

\begin{enumerate}
    \item Use quantum and classical compute to calculate the mitigated observable for the target circuit: $\left<O\right>_{\mathbb{M}}$.

    \item Use classical compute to calculate the ideal expectation values for the ensemble circuits $\mathcal{C} \in \mathscr{B}$: ${\rm Tr}\left[\rho\mathcal{C}^\dagger(O)\right]$.

    \item Use quantum and classical compute to calculate the mitigated expectation values for the ensemble circuits $\mathcal{C} \in \mathscr{B}$: ${\rm Tr}\left[\rho\tilde{\mathcal{C}}^\dagger(O)\right]_\mathbb{M}$.

    \item Compute the boosted expectation value:
    \begin{align}
        \nonumber\left<O\right>_{\mathbb{B}} = \left<O\right>_{\mathbb{M}} + \sum_{\mathcal{C} \in \mathscr{B}} g(\mathcal{C})\left({\rm Tr}\left[\rho\mathcal{C}^\dagger(O)\right] - {\rm Tr}\left[\rho\tilde{\mathcal{C}}^\dagger(O)\right]_\mathbb{M}\right).
    \end{align}
\end{enumerate}

\subsection{Why Boosted Error Mitigation Works}

To understand why the BEM procedure works, we first note the trivial fact that
\begin{align}
    \left<O\right> = \left<O\right> + \left<O\right>_{\mathbb{M}} - \left<O\right>_{\mathbb{M}} = \left<O\right>_{\mathbb{M}} + \delta_{\mathbb{M}} ,
\end{align}
where we have defined the mitigation bias error $\delta_{\mathbb{M}} = \left<O\right> - \left<O\right>_{\mathbb{M}}$. Assuming that property 2 of the BEM requirements holds for our chosen mitigation procedure, we can express the bias as
\begin{align}
    \nonumber\delta_{\mathbb{M}} &= \sum_{\mathcal{C} \in \mathscr{E}} g(\mathcal{C}){\rm Tr}\left[\rho\mathcal{C}^\dagger(O)\right] - \sum_{\mathcal{C} \in \mathscr{E}} g(\mathcal{C}){\rm Tr}\left[\rho\tilde{\mathcal{C}}^\dagger(O)\right]_\mathbb{M} \\
    &= \sum_{\mathcal{C} \in \mathscr{E}} g(\mathcal{C}) \delta_\mathcal{C},
\end{align}
where $\delta_\mathcal{C}$ is the bias error remaining for the mitigated expectation value of the ensemble circuit $\mathcal{C}$. 

Now, let $\delta_{\mathbb{M}}^{B}$ be a partial summation of $\delta_{\mathbb{M}}$ over a subset of circuits $\mathscr{B}\subset\mathscr{E}$, chosen to satisfy the hierarchy defined in property 4 of the BEM requirements, i.e.~
\begin{align}
    \delta_{\mathbb{M}}^{B} = \sum_{\mathcal{C} \in \mathscr{B}} g(\mathcal{C}) \delta_\mathcal{C}.
\end{align}
This is exactly the quantity that steps 2 and 3 of the BEM protocol estimate, which becomes clear when we see that we can rewrite the

boosted expectation value produced by BEM as
\begin{align}
    \left<O\right>_{\mathbb{B}} = \left<O\right>_{\mathbb{M}} + \delta_{\mathbb{M}}^{B}.
\end{align}
The remaining bias error \emph{after} boosting is given by
\begin{align}
     \delta_{\mathbb{B}} = \left<O\right> - \left<O\right>_{\mathbb{B}} = \delta_{\mathbb{M}} - \delta_{\mathbb{M}}^{B}, 
\end{align}
and if requirement 4 of the BEM protocol is met, we have that $\abs{\delta_{\mathbb{M}} - \delta_{\mathbb{M}}^{B}} < \abs{\delta_{\mathbb{M}}}$ since the circuits in the ensemble subset $\mathscr{B}$ have the largest coefficients. Thus, the boosted expectation value is a more accurate approximation to the ideal expectation value than the original mitigated result. In the limit where the entire set is used for boosting, i.e.~$\mathscr{B} = \mathscr{E}$, then the boosting procedure trivially returns the ideal expectation value as calculated by classical simulation of the series expansion. While needlessly inefficient, this guarantees that BEM is asymptotically bias free. 

\subsection{Comparing Boosted Error Mitigation to Classical Simulation}

The output of a BEM protocol is naturally compared to the classical estimation of the ideal expectation value using the series representation
\begin{align}
    \left<O\right>_{\mathbb{S}} =  \sum_{\mathcal{C} \in \mathscr{B}} g(\mathcal{C}){\rm Tr}\left[\rho\mathcal{C}^\dagger(O)\right],
\end{align}
which has the same cost in terms of the number of circuits to be executed. For any reasonable mitigation procedure, the boosted result is more accurate than this direct series estimation. To see this, consider the remaining bias for each procedure coming from the circuits not in the set $\mathscr{B}$, which we denote by the set $\lnot\mathscr{B}$, given by
\begin{align}
     &\nonumber\delta_{\mathbb{B}} = \sum_{\mathcal{C} \in \lnot\mathscr{B}} g(\mathcal{C}) \delta_\mathcal{C}, \\
     &\delta_{\mathbb{S}} =  \sum_{\mathcal{C} \in \lnot\mathscr{B}} g(\mathcal{C}){\rm Tr}\left[\rho\mathcal{C}^\dagger(O)\right].
\end{align}
For any circuit, a reasonable mitigation procedure will have bias smaller than the ideal expectation value, such that $\abs{\delta_\mathcal{C}} < \abs{{\rm Tr}\left[\rho\mathcal{C}^\dagger(O)\right]}$. Thus, for a series with approximately monotonic convergence, we have that $\abs{\delta_{\mathbb{B}}} < \abs{\delta_{\mathbb{S}}}$, and BEM is more accurate than direct simulation, once the set $\mathscr{B}$ is big enough to smooth out any initial nonmonotonic behavior in the early series terms.

Within the Pauli twirling limit, we can more concretely elucidate the advantage over CPT classical simulation for QuEPP. The expectation value of each Clifford circuit in the ensemble will be scaled by a factor $\eta_{i,k}$ that represents the impact of noise. With this we can write the boosted expectation value as
\begin{align}
    \left<O\right>^{K_T}_{\mathbb{B}} &= \left<O\right>_{\mathbb{M}}+\sum_{k=0}^{K_T} \sum_{i=1}^{N_k} g(i,k)\left(1 - \frac{\eta_{i,k}}{\eta}\right){\rm Tr}\left[\rho\mathcal{C}_{i,k}^\dagger(O)\right],
\end{align}
and the remaining bias in the QuEPP-boosted expectation value is given by
\begin{align}
    \nonumber\delta^{K_T}_{\mathbb{B}} &= \left<O\right> - \left<O\right>^{K_T}_{\mathbb{B}} \\ &=\sum_{k=K_T+1}^K \sum_{i=1}^{N_k} g(i,k)\left(1 - \frac{\eta_{i,k}}{\eta}\right){\rm Tr}\left[\rho\mathcal{C}_{i,k}^\dagger(O)\right].
\end{align}
Comparing this to the remaining bias for a classical CPT simulation truncated to order $K_T$
\begin{align}
    \delta^{K_T}_{\mathbb{S}} = \sum_{k=K_T+1}^K \sum_{i=1}^{N_k} g(i,k){\rm Tr}\left[\rho\mathcal{C}_{i,k}^\dagger(O)\right],
\end{align}
we have that for reasonably well-behaved series that have approximate monotonic convergence, $\abs{1 - \eta_{i,k}/\eta} < 1$ implies that $\abs{\delta^{K_T}_{\mathbb{B}}} < \abs{\delta^{K_T}_{\mathbb{S}}}$. Thus, under the assumption that our initial error mitigation protocol does a reasonable job, the remaining bias from QuEPP will be less than that from classical simulation of the CPT series, as our experimental demonstrations have shown. Guaranteeing that $\abs{1 - \eta_{i,k}/\eta} < 1$ for all circuits depends strongly on the choice of $\eta$, but in practice we find this does not seem to be an issue, and small violations on a small subset of circuits is not sufficient to impact the accuracy of QuEPP.

\section{QuEPP Variance Scaling}

In this section we derive analytical estimates for how the variance in a QuEPP boosted expectation value scales with the number of circuit executions. As a reminder, the purpose of QuEPP, and any QEM protocol, is to reduce bias, which is one component in the error of the expectation value. The other is statistical variance, which unfortunately for most QEM protocols increases.

To quantify this increased variance, we assume that the classical executions in the CPT calculations used in QuEPP have zero uncertainty. The quantum executions in QuEPP are used to estimate the bias term, which for Pauli stochastic noise can be written as
\begin{align}
    &\nonumber \delta_{\mathbb{M}}^{K_T} = \sum_{k=0}^{K_T} \sum_{i=1}^{N_k} g(i,k)\left(1 - \frac{\eta_{i,k}}{\eta}\right){\rm Tr}\left[\rho\mathcal{C}_{i,k}^\dagger(O)\right] \\
    &= \sum_{k=0}^{K_T} \sum_{i=1}^{N_k} g(i,k)\left({\rm Tr}\left[\rho\mathcal{C}_{i,k}^\dagger(O)\right] - \frac{1}{\eta}{\rm Tr}\left[\rho\tilde{\mathcal{C}}_{i,k}^\dagger(O)\right]\right),
\end{align}
which ultimately depends on the expectation value of the noisy quantum circuit. Since this noisy expectation value is the mean of a binomial random variable with possible outcomes $\pm 1$, it is straightforward to calculate the variance of $\delta_{\mathbb{M}}^{K_T}$, which is nothing but the standard error of the mean. For $N$ circuit executions, this is given by
\begin{align}
    \sigma^2\left[\delta_{\mathbb{M}}^{K_T}\right] = \sum_{k=0}^{K_T} \sum_{i=1}^{N_k} \abs{g(i,k)}^2\frac{1}{\eta^2 N}\left(1 - \eta_{i,k}^2\right),
\end{align}
which shows the expected $1/N$ scaling. To make this expression a bit more tractable, as $\abs{\eta_{i,k}} < 1$ we can simply upper bound $1 - \eta_{i,k}^2 < 1$, to obtain
\begin{align}
    \sigma^2\left[\delta_{\mathbb{M}}^{K_T}\right] \leq \sum_{k=0}^{K_T} \sum_{i=1}^{N_k} \abs{g(i,k)}^2\frac{1}{\eta^2 N} = \frac{1}{\eta^2 N} P_{K_T},
\end{align}
where we have lumped the summation of the squared coefficients into $P_{K_T}$. We note that $P_{K_T}$ for a given circuit can be calculated exactly from the CPT ensemble expansion, and is always smaller than one.

Finally, to compare to the resource scaling used in Probabilistic Error Cancellation \cite{Berg22} we observe that rescaling by a parameter $\eta$ can be thought of as error mitigation assuming a depolarizing error channel. Then, we can express $\eta$ in terms of the $\gamma$ resource parameter of PEC as $\eta^2 = 1/\gamma$, which gives us our final expression for the variance
\begin{align}
    \sigma^2\left[\delta_{\mathbb{M}}^{K_T}\right] \leq \frac{\gamma}{N} P_{K_T}.
\end{align}

\section{Bounding the Remaining Bias}
\label{sec:bias}

In this section we will discuss ways to upper bound the remaining bias error after applying the QuEPP protocol to order $K_T$, given by the expression
\begin{align}
    \delta^{K_T}_{\mathbb{B}} = \sum_{k=K_T+1}^K \sum_{i=1}^{N_k} g(i,k)\left(1 - \frac{\eta_{i,k}}{\eta}\right){\rm Tr}\left[\rho\mathcal{C}_{i,k}^\dagger(O)\right].
\end{align}
Our first approach follows the methodology of \cite{Lerch24}. We begin by using the triangle inequality to express the bias as
\begin{align}
    \abs{\delta^{K_T+1}_{\mathbb{B}}} \leq \sum_{k=K_T+1}^K \sum_{i=1}^{N_k} 
\abs{g(i,k)}\abs{1 - \frac{\eta_{i,k}}{\eta}},
\end{align}
where we have used the fact that the expectation value of a Pauli observable propagated backwards through a Clifford circuit with a state $\rho$ that is a Clifford eigenstate can only take the values $\pm 1$ or $0$. Using the expression for $g(i,k)$ in the main text, we have that $\abs{g(i,k)}\leq \sin^k(\theta^*)$ where $\theta^*$ is the largest rotation angle of the non-Clifford gates in our target circuit. Similarly, we can define $\eta^*$ that maximizes the expression $\abs{1 - \eta_{i,k}/\eta}$. Note that this necessarily makes this upper-bound approximate, as we can only find $\eta^*$ from the subset of ensemble circuits for which we have measured $\eta_{i,k}$. However, as upper-bounding this way can lead to a loose upper bound, we do not think this makes much difference in practice.

With these identifications, using combinatorial identities we can arrive at the upper bound
\begin{align}
    \nonumber\abs{\delta^{K_T}_{\mathbb{B}}} &\leq \abs{1 - \frac{\eta^*}{\eta}}\sum_{k=K_T+1}^K {K \choose k} \sin^k(\theta^*) \\ &\leq \abs{1 - \frac{\eta^*}{\eta}}\left(\frac{eK\sin\theta^*}{K_T+1}\right)^{K_T+1},
\end{align}
where $e$ is Euler's number. The second line of this upper bound is only valid when $\sin\theta^* \leq (K_T+1)/K$, but the first line is always valid and can be calculated numerically. In this expression, to get an upper bound we have assumed that all possible circuits of a given order $k$ both form part of the ensemble and have nonzero expectation value. For a random target circuit, on average only a $1/2^n$ fraction of ensemble circuits will have non-zero expectation value. In practice, this discrepancy makes the upper bound loose, and so we will now consider an alternative approach.

To start, we notice that $\delta^{K_T}_{\mathbb{B}}$ can be written as the difference of the ideal CPT series and the mitigated CPT series for orders $K_T+1$ to $K$. The ideal CPT series from orders $K_T+1$ to $K$ is equivalent to the ideal value minus the ideal CPT series from $0$ to $K_T$, that is
\begin{align}
    &\nonumber \sum_{k=K_T+1}^K \sum_{i=1}^{N_k} g(i,k){\rm Tr}\left[\rho\mathcal{C}_{i,k}^\dagger(O)\right] \\ &= \left<O\right> - \sum_{k=0}^{K_T} \sum_{i=1}^{N_k} g(i,k){\rm Tr}\left[\rho\mathcal{C}_{i,k}^\dagger(O)\right].
\end{align}
Similarly for the mitigated expression we can write
\begin{align}
    &\nonumber \sum_{k=K_T+1}^K \sum_{i=1}^{N_k} \frac{\eta_{i,k}}{\eta} g(i,k){\rm Tr}\left[\rho\mathcal{C}_{i,k}^\dagger(O)\right] \\ &= \left<O\right>_{\mathbb{M}} - \sum_{k=0}^{K_T} \sum_{i=1}^{N_k}  \frac{\eta_{i,k}}{\eta} g(i,k){\rm Tr}\left[\rho\mathcal{C}_{i,k}^\dagger(O)\right],
\end{align}
and overall arrive at
\begin{align}
    \delta^{K_T}_{\mathbb{B}}&= \left<O\right> - \left<O\right>_{\mathbb{M}}\\ & - \nonumber\sum_{k=0}^{K_T} \sum_{i=1}^{N_k} g(i,k)\left(1 - \frac{\eta_{i,k}}{\eta}\right){\rm Tr}\left[\rho\mathcal{C}_{i,k}^\dagger(O)\right].
\end{align}
We note that this is really just the trivial fact that the remaining bias is the difference between the ideal value and the boosted value, i.e.~
\begin{align}
    \delta^{K_T}_{\mathbb{B}} =  \left<O\right> - \left<O\right>^{K_T}_{\mathbb{B}} = \left<O\right> - \left<O\right>_{\mathbb{M}} - \delta^{K_T}_{\mathbb{M}}.
\end{align}

Next, we make the reasonable assumption that $K_T$ is large enough and the CPT series sufficiently well-behaved that increasing $K_T$ reduces $\delta^{K_T}_{\mathbb{B}}$ monotonically, with $\left<O\right> - \left<O\right>_{\mathbb{M}}$ and $\delta^{K_T}_{\mathbb{B}}$ both having the same sign. In this case, we have that
\begin{align}
    \abs{\delta^{K_T}_{\mathbb{B}}}  \leq  \abs{\left<O\right> - \left<O\right>_{\mathbb{M}}} - \abs{\delta^{K_T}_{\mathbb{M}}}.
\end{align}
While this may look like the triangle inequality, the minus sign between the two terms on the right-hand side relies on the fact that $\delta^{K_T}_{\mathbb{B}}$ is monotonically decreasing as a function of $K_T$, in which case it is safe to assume that $\delta^{K_T}_{\mathbb{M}}$ will have the same sign as $\left<O\right> - \left<O\right>_{\mathbb{M}}$.

However, this expression still contains the inaccessible quantity $\left<O\right>$, which for non-Clifford target circuits can range anywhere between $\pm 1$. To get around this, we factor out the $\left<O\right>_{\mathbb{M}}$ and use the results of Ref.~\cite{Govia25} to obtain
\begin{align}
    \nonumber\abs{\delta^{K_T}_{\mathbb{B}}}  &\leq  \abs{\frac{\left<O\right>}{\left<O\right>_{\mathbb{M}}} - 1}\left<O\right>_{\mathbb{M}} - \abs{\delta^{K_T}_{\mathbb{M}}} \\
    &\lessapprox  \abs{\frac{\eta}{\eta'} - 1}\left<O\right>_{\mathbb{M}} - \abs{\delta^{K_T}_{\mathbb{M}}}, \label{eqn:upperbound}
\end{align}
where $\eta'$ maximizes the expression $\abs{1 - \eta/\eta_{i,k}}$ for the $\eta_{i,k}$ of the measured ensemble circuits. The second line in the above expression upper bounds the distance between the ideal and initial mitigated expectation value of the target circuit by the Clifford circuit (from the measured ensemble) whose mitigated expectation value is furthest from its ideal value. As shown in Ref.~\cite{Govia25} this is not a strict upper bound, but it is a good heuristic that functions as an upper bound except for pathological cases. In fact, replacing the worst case $\abs{\eta/\eta' - 1}$ in Eq.~\eqref{eqn:upperbound} with the average over the ensemble  $\abs{\eta/\bar{\eta} - 1}$ is likely to give a closer estimate to the remaining bias for most target circuits \cite{Merkel25}.

\section{Choice of Rescaling Parameter Used in QuEPP}\label{sec:bound}

The optimal choice of rescaling parameter $\eta$ is such that it minimizes the remaining bias
\begin{align}
    \delta^{K_T}_{\mathbb{B}} = \sum_{k=K_T+1}^K \sum_{i=1}^{N_k} g(i,k)\left(1 - \frac{\eta_{i,k}}{\eta}\right){\rm Tr}\left[\rho\mathcal{C}_{i,k}^\dagger(O)\right].
\end{align}
However, calculating this is impossible without knowing the $g(i,k)$ or $\eta_{i,k}$ of this sum, and as such we are forced to use approximate strategies. One approach is to rewrite the remaining bias as was done in the previous section
\begin{align}
    \delta^{K_T}_{\mathbb{B}}&= \left<O\right> - \left<O\right>_{\mathbb{M}}\\ & - \nonumber\sum_{k=0}^{K_T} \sum_{i=1}^{N_k} g(i,k)\left(1 - \frac{\eta_{i,k}}{\eta}\right){\rm Tr}\left[\rho\mathcal{C}_{i,k}^\dagger(O)\right],
\end{align}
and then set the term on the second line exactly to zero by choosing
\begin{align}
    \eta = \frac{\sum_{k=0}^{K_T} \sum_{i=1}^{N_k} g(i,k)\eta_{i,k}{\rm Tr}\left[\rho\mathcal{C}_{i,k}^\dagger(O)\right]}{\sum_{k=0}^{K_T} \sum_{i=1}^{N_k} g(i,k){\rm Tr}\left[\rho\mathcal{C}_{i,k}^\dagger(O)\right]}.
\end{align}
This expression can be computed using the fact that we have executed all relevant Clifford circuits up to order $K_T$ and have the full set of $g(i,k)$ and $\eta_{i,k}$ needed. This choice of $\eta$ is a weighted average of the per-circuit rescaling factors $\eta_{i,k}$, weighted by their (signed) circuit coefficient. 

The tradeoff of this choice of $\eta$ is that it has an unknown impact on the term in the first line of the remaining bias, $\left<O\right> - \left<O\right>_{\mathbb{M}}$. However, as $K_T$ increases this choice for $\eta$ approaches the exact rescaling factor needed to correct the noisy expectation value, and as such  $\abs{\left<O\right> - \left<O\right>_{\mathbb{M}}} \rightarrow 0$ as $K_T$ increases. Thus, when sampling a large amount of the ensemble is possible, this choice for $\eta$ is a reliable one. Note that when the circuit coefficients are all close to equal, this choice of $\eta$ can be very close to the mean of the $\eta_{i,k}$ set.

In situations where sampling is limited, such as when Monte Carlo sampling is deployed for circuits with many non-Clifford gates, the weighted average choice for $\eta$ can be problematic due to limited data and small coefficient sizes. Instead, returning to the original expression for the remaining bias, we can aim to minimize (on average) each term in the sum by minimizing
\begin{align}
    \sum_{k=K_T+1}^K \sum_{i=1}^{N_k} \abs{1 - \frac{\eta_{i,k}}{\eta}}.
\end{align}
However, since we do not know $\eta_{i,k}$ for this part of the sum, we assume that the $\eta_{i,k}$ we do know are representative of the full ensemble and solve the minimization problem
\begin{align}
    \min_{\eta}\sum_{k=0}^{K_T} \sum_{i=1}^{N_k} \abs{1 - \frac{\eta_{i,k}}{\eta}}.
\end{align}
This is \emph{not} minimized by the median of the $\eta_{i,k}$ ensemble, which would be the solution if the argument above was $\abs{\eta_{i,k}-\eta}$, but instead by $\eta$ that satisfies
\begin{align}
    \sum_{\eta_{i,k} < \eta} \eta_{i,k} = \sum_{\eta_{i,k} > \eta}\eta_{i,k},
\end{align}
which will generally be larger than the median or mean. Intuitively, this follows from the fact that the remaining bias for a circuit with $\eta_{i,k} > \eta$ can grow without bound, while the bias for a circuit with $\eta_{i,k} < \eta$ is upper bounded by one. Thus, it is better to rescale with less effective circuit noise by choosing a larger $\eta$. 

\section{Generalization of QuEPP beyond order-based method}

In the main text and in Sec.~\ref{sec:bound}, we presented QuEPP in its order-based formulation.  In this section, we provide a description of QuEPP in a more general setting.  
As in the paper, we use $\tilde{\mathcal{U}}$ ($\tilde{\mathcal{C}})$ to denote the noisy version of the ideal target (Clifford) circuit $\mathcal{U}$ ($\mathcal{C}$).

The generalized steps are as follows:

\begin{enumerate}
    \item Using classical compute, sample a collection $\mathcal{B}$ of unique CPT ensemble circuits from the desired distribution $\mathcal{D}$, under the restriction that each circuit in $\mathcal{B}$ must have a non-zero ideal expectation value.  Denote each sampled CPT ensemble circuit in $\mathcal{B}$ by its index and order: $(i,k)$.  Calculate the ideal expectation values for each circuit in $\mathcal{B}$: ${\rm Tr}\left[\rho\mathcal{C}_{i,k}^\dagger(O)\right]$.
    And the classical estimate given the collection $\mathcal{B}$:
    \begin{align}
        \left<O\right>^{\mathcal{B}} = \sum_{(i,k) \in \mathcal{B}} g(i,k)\,{\rm Tr}\left[\rho{\mathcal{C}}_{i,k}^\dagger(O)\right].
    \end{align}

    \item On the quantum computer estimate the noisy expectation values of: 
    
    \begin{enumerate}
        \item The target circuit: $\left<O\right>_{\rm noisy} = {\rm Tr}\left[\rho\tilde{\mathcal{U}}^\dagger(O)\right]$.
        \item The Clifford circuits in $\mathcal{B}$: ${\rm Tr}\left[\rho\tilde{\mathcal{C}}_{i,k}^\dagger(O)\right]$. The noisy estimate given the collection $\mathcal{B}$ of the target circuit is 
        \begin{align}
            \left<O\right>^{\mathcal{B}}_{\rm noisy} = \sum_{(i,k) \in \mathcal{B}} g(i,k)\,{\rm Tr}\left[\rho\tilde{\mathcal{C}}_{i,k}^\dagger(O)\right].
        \end{align}
        
    \end{enumerate}

    \item Subtract the noisy estimate of the expectation values given $\mathcal{B}$ from the noisy expectation value of the target circuit. This gives us a noisy estimate of the paths we did not compute classically
    \begin{align}
        \nonumber\left<O\right>^{\lnot\mathcal{B}}_{\rm noisy} = \left<O\right>_{\rm noisy} -  \left<O\right>^{\mathcal{B}}_{\rm noisy}.
    \end{align}
    
    \item From the distribution of per-circuit rescaling factors $\eta_{i,k} = {\rm Tr}\left[\rho\tilde{\mathcal{C}}_{i,k}^\dagger(O)\right]/{\rm Tr}\left[\rho\mathcal{C}_{i,k}^\dagger(O)\right]$ of the CPT ensemble, systematically determine a global rescaling factor $\eta$.

    \item Compute the enhanced expectation value
    \begin{align}
        \nonumber\left<O\right>^{\mathcal{B}}_{\mathbb{M}} = 
        \left<O\right>^{\mathcal{B}} +  \left<O\right>^{\lnot{\mathcal{B}}}_{\rm noisy} /\eta   
    \end{align}

Given this generalized definition, we introduce three specific variants of the method:

\begin{itemize}
    \item The order-based method, as presented in the main text, is equivalent to the above procedure where $\mathcal{B} = \{\, (i,k) \mid k \le K_T \,\}$.
    \item The coefficient-based method is similar to the order-based method, but truncates on a small coefficient rather than the order.  Specifically, given a threshold $\epsilon$, $\mathcal{B} = \{\, (i,k) \mid \left| g(i,k) \right| \ge \epsilon \,\}$.
    \item The Monte Carlo method, where $\mathcal{B}$ is formed by sampling Pauli paths, without replacement, from some desired distribution $\mathcal{D}$ of Pauli paths.  For instance, one suitable distribution is that where the probability of sampling path $(i,k)$ is proportional to its absolute coefficient, $\left| g(i,k) \right|$.
\end{itemize}

The Monte Carlo method is intriguing because, when it is tractibile, it allows the Pauli paths to be sampled from the set of paths that contribute to the ideal expectation value, in an unbiased way, in proportion to their contribution to this value.  However, Pauli paths with a non-zero ideal expectation value can be very rare (up to exponentially rare for a random circuit), so sampling from this distribution can be challenging for classical compute.  A method for sampling from a related distribution is described in the following section.

On the other hand, the order-based and coefficient-based methods can be efficiently obtained through a depth-first search, as the order of a Pauli path can only increase as the depth grows, while the coefficient of a Pauli path can only decrease as depth grows.

\section{Monte Carlo method of searching for paths}

As in the paper, we start with a target circuit $\mathcal{U}$ that contains all Clifford operations except for $K$ Pauli rotation gates, as well as a Pauli observable $O$.
Ideally, the Monte Carlo method would sample from the distribution where the path $(i,k)$ is drawn with probability proportional to $\left| g(i,k) \right|$.
Our goal is to generate samples as quickly as possible, and we only care about a sparse subset of them: those that result in non-zero amplitude when an observable of interest is propagated through the circuit and projected onto the initial state $\ket{0}$.

Let us start with the definition of the coefficient in the paper:
\begin{equation}
    g(i,k) = \prod_{j=1}^k \left( s_{i,j} \sin{\theta_{i,j}} \right) \prod_{j=k+1}^{K}\left(\cos{\theta_{i,j}}\right)^{s_{i,j}} . \label{eqn:CPTcoeff2}
\end{equation}
Recall that the parameter $s_{i,k}$ is zero (one) if the back-propagated observable commutes (anti-commutes) with the generator of the Pauli rotation gate at a given layer.  In the above product, the labels $j = 1, \ldots, k$ denote non-Clifford instructions at which the $\sin$ branch was taken, and the labels $j=k+1,\dots,K$ denote non-Clifford instructions where either the $\cos$ branch was taken ($s_{i,k}=1$ case), or else no factor is necessary because the generator of Pauli rotation commutes with the back-propagated observable ($s_{i,k}=0$ case).  Here, we have explicitly placed a sub-factor of $s_{i,j}$ within each $\sin$-branch factor to indicate that if a history attempts to take the $\sin$ branch at a location where the back-propagated observable commutes with the generator of Pauli rotation, then it should be eliminated from consideration.

The Monte Carlo procedure we use is equivalent to drawing a path using the following procedure.  Back-propagate the observable until a non-Clifford is reached.  If the observable commutes with the generator of the Pauli rotation, then continue back-propagation.  Otherwise, take the $\cos$ branch with probability $|\cos\theta| / \left[ |\cos\theta| + |\sin\theta| \right]$ and the $\sin$ branch otherwise.  Repeat this procedure until the beginning of the circuit is reached.  At this point, test if the back-propagated observable commutes with $Z^{\otimes n}$.  If it does, then it is a path which contributes to the observable's expectation value.  We then add the path to the QuEPP ensemble if it is not already represented there.

The above procedure actually samples a distribution $\mathcal{\tilde D}$ that differs from the ideal distribution $\mathcal{D}$, where each path $(i,k)$ is sampled in proportion to $\left| g(i,k) \right|$.  The problem is that, in the $s_{i,k}=1$ case, the absolute coefficient was rescaled by a factor of $1 / \left[ |\cos\theta| + |\sin\theta| \right]$, but the $s_{i,k}=0$ points did not receive that rescaling.  Thus, the above-described simulation is biased toward paths with more points that commute with the back-propagated Pauli, up to a factor of $\sqrt 2$ per point (reached when $\theta=\pi/4$).  In this work, we have used the distribution $\tilde{\mathcal{D}}$, but there are ways one might adjust the procedure in order to sample from $\mathcal
{D}$ instead.  The first method would be to perform post-selection.  In this case, modify the above procedure such that when $s_{i,k}=0$, continue the path with probability $1 / \left[ |\cos\theta| + |\sin\theta| \right]$, otherwise abort the path.
A more sophisticated approach would be to perform a population Monte Carlo scheme in order to sample from the distribution $\mathcal{D}$ directly.  We leave investigation of this possibility for future work.

\end{enumerate}

\clearpage{}

\clearpage{}

\clearpage{}

\end{document}